\documentclass[12pt]{article}

\usepackage{epsfig}			

\newcommand{\nb}[0]{{N_{B}}}
\newcommand{\ns}[0]{{N_{S}}}
\newcommand{\nbar}[0]{\overline{n}}
\newcommand{\bitz}[0]{Z_{0, \nb -1 }}
\newcommand{\sx}[0]{\sigma_x}
\newcommand{\sy}[0]{\sigma_y}
\newcommand{\sz}[0]{\sigma_z}
\newcommand{\cnotyes}[2]{\sx(#2)^{n(#1)}}
\newcommand{\cnotno}[2]{\sx(#2)^{\nbar(#1)}}
\newcommand{\va}[0]{{\vec{a}}}
\newcommand{\vb}[0]{{\vec{b}}}
\newcommand{\caln}[0]{{\cal N}}

\newcommand{\ket}[1]{|#1\rangle}

\newcommand{\beq}{\begin{equation}}  
\newcommand{\eeq}{\end{equation}}
\newcommand{\beqa}{\begin{eqnarray*}}  
\newcommand{\eeqa}{\end{eqnarray*}}
\newcommand{\rarrow}[0]{\rightarrow}

\newcommand{\enote}[1]{\cite{#1}}    
\newcommand{\eqlabel}[1]{\renewcommand{\theequation}{#1}}

\begin{document}
\title{A Rudimentary Quantum Compiler (2cnd Ed.)}

\author{Robert R. Tucci\\
        P.O. Box 226\\ 
        Bedford,  MA   01730\\
        tucci@ar-tiste.com}

\date{ \today} 

\maketitle

\vskip2cm
\section*{Abstract}
We present a new algorithm for reducing an arbitrary
unitary matrix U into a sequence of elementary operations 
(operations such as controlled-nots and qubit rotations). Such a sequence of
operations can be used to manipulate an array of quantum bits (i.e.,
a quantum computer). Our algorithm
 applies recursively a mathematical
technique called the CS Decomposition
to build a binary tree of matrices 
whose product, in some order, equals the original matrix U.
We show that the Fast Fourier Transform (FFT) algorithm is a 
special case of our algorithm.
We report on a C++ program 
called ``Qubiter" that implements the ideas of this paper. 
Qubiter(PATENT PENDING) source code
is publicly available.

\newpage
\section*{1. Introduction}
\subsection*{1(a) Previous Work}
\mbox{}\indent	
In classical computation and digital electronics, one 
deals with sequences of elementary operations
(operations such as AND, OR and NOT).
These sequences are used to manipulate an array of classical bits.
The operations are elementary in the sense that they 
act on only a few bits (1 or 2) at a time.
 Henceforth, we will sometimes refer to sequences as products and to 
operations as operators, matrices, instructions, steps or gates.
Furthermore, we will  abbreviate 
the phrase ``sequence of elementary operations" by ``SEO".
In quantum computation\enote{QC}, one also deals with SEOs 
(with operations such as  
controlled-nots and qubit rotations), but for manipulating 
quantum bits (qubits) instead of classical bits.
Quantum SEOs are often represented graphically by qubit circuits.

In quantum computation, one
often knows the unitary operator $U$ that describes the evolution 
of an array of qubits. One must then 
find a way to reduce $U$ into a
SEO. 
In this paper, we present a new algorithm 
for accomplishing this task. We also report on a C++ program called
``Qubiter" that implements our algorithm. 
 We call Qubiter
 a ``quantum compiler" because, like a  
classical compiler, it produces a SEO for manipulating bits.
Qubiter(PATENT PENDING)
source code is publicly
available at www.ar-tiste.com/qubiter.html.

Our algorithm can be applied to any
unitary operator $U$. 

It is useful to define certain unitary operators $U_{\nb}$
for all $\nb \in \{1, 2, 3, ...\}$, where
$U_{\nb}$ is a $2^\nb \times 2^\nb$ matrix and 
$\nb$ is the number of bits.
Some $U_\nb$ are known to be expressible as 
a SEO whose length (i.e., whose number of elementary operations) 
is a polynomial in $\nb$. Two examples are the $\nb$ bit
 Hadamard Transform (HT) matrix
and the $\nb$ bit  Discrete Fourier Transform (DFT) matrix. 
The HT matrix is known to be expressible as a SEO of length Order($\nb$). 
The DFT matrix is known to be
expressible (using the FFT algorithm of \enote{Knuth}-\enote{Grif}) as a SEO of length Order($\nb^2$). Our algorithm 
achieves both of these SEO-length benchmarks. 
Even better, the SEO often called the ``quantum FFT algorithm"
is exactly reproduced by our algorithm.

We believe our algorithm yields 
short SEOs for many kinds of unitary operators other
than the HT and DFT ones. We do not believe that our present 
algorithm yields the
shortest possible SEO for every unitary operator. However, 
we do believe that it is possible
to come very close to achieving this goal
by introducing further optimizations into the algorithm. 
Future papers will report on our progress in finding such optimizations.

Previous workers \enote{CastOfThousands} have described another 
algorithm for reducing a unitary operator into a SEO. 
Like ours, their algorithm can be applied to any
unitary operator $U$.  However, 
it is very unlikely  that their algorithm
will be efficient in producing short SEOs unless
further optimizations are added to it. And
such optimizations, if they exist, have not been specified by anyone.
Furthermore, as far as we know, there is no publicly available software 
that implements their algorithm.
Our algorithm is significantly
different from theirs. Theirs is based on a mathematical technique
described in Refs.\enote{Murn}-\enote{Reck}, whereas ours is based on
a mathematical technique called the CS Decomposition
(CSD)\enote{Stew}-\enote{Bai}
to be described later.

Quantum Bayesian (QB) Nets\enote{Tucci95}-\enote{QFog} 
are a method of modeling quantum
systems graphically in terms of network diagrams. In a 
companion paper\enote{Tucci98b}, we  show how to apply the
results of this paper to QB nets.

\subsection*{1(b) CS Decomposition}
\mbox{}\indent	
As mentioned earlier, our algorithm utilizes a mathematical technique
called the CS Decomposition
(CSD)\enote{Stew}-\enote{Bai}. The C and S stand for ``cosine" and ``sine", respectively.
Next we will state
the special case of the CSD Theorem
that arises in our algorithm. 

Suppose that $U$ is an $N\times N$ unitary matrix, where 
$N$ is an even number.
Then the CSD Theorem states that one can always 
express $U$ in the form 

\beq
U = 
\left [
\begin{array}{cc}
L_0 &  0 \\
0   &  L_1
\end{array}
\right ]
D
\left [
\begin{array}{cc}
R_0 &  0 \\
0   &  R_1
\end{array}
\right ]
\;,
\eqlabel{1b.1}\eeq
where the left and right side matrices $L_0, L_1, R_0, R_1$ are 
$\frac{N}{2}\times \frac{N}{2}$  unitary matrices and

\beq
D = 
\left [
\begin{array}{cc}
D_{00} &  D_{01} \\
D_{10}  &  D_{11}
\end{array}
\right ]
\;,
\eqlabel{1b.2a}\eeq

\beq
D_{00} = D_{11} = diag(C_1, C_2, \ldots, C_{\frac{N}{2}})
\;,
\eqlabel{1b.2b}\eeq

\beq
D_{01} = diag(S_1, S_2, \ldots, S_{\frac{N}{2}})
\;,
\eqlabel{1b.2c}\eeq

\beq
D_{10} = - D_{01}
\;.
\eqlabel{1b.2d}\eeq
For all $i \in \{ 1, 2, \ldots , \frac{N}{2} \}  $,
 $C_i = \cos\theta_i$ and $S_i = \sin\theta_i$
for some angle $\theta_i$. 
Given any CSD of $U$,
it is easy to find (see Appendix A) 
another CSD of $U$ for which the angles $\theta_i$ are in non-decreasing
order and they are contained in the interval $[0, 90^0]$.
Henceforth, we will assume that the angles $\theta_i$ are so ordered and
in this range.
We will use the term 
{\it D matrix} to refer to any matrix that satisfies Eqs.(1b.2).
If one partitions $U$ into four blocks $U_{ij}$ 
of size $\frac{N}{2}\times \frac{N}{2}$ ,
then 

\beq
U_{ij} = L_i D_{ij} R_j
\;,
\eqlabel{1b.3}\eeq
for $i, j \in \{0, 1\}$. Thus, $D_{ij}$ gives the singular values\enote{Noble}
of $U_{ij}$.

 More general versions of the CSD Theorem allow for the
possibility that we
partition $U$ into 4 blocks
of unequal size. 

Note that if $U$ were a general (not necessarily unitary) matrix, then
the four blocks $U_{ij}$ would be unrelated. 
Then to find the singular values of the four blocks 
$U_{ij}$ would require
 eight unitary matrices (two for each block), 
 instead of the four $L_i, R_j$.
 This double use of the $L_i, R_j$ is a key property 
 of the CSD.

\subsection*{1(c) Bird's Eye View of Algorithm}
\mbox{}\indent	
Our algorithm is described in detail in subsequent sections.
Here we will only give a bird's eye view of it.

		\begin{center}
			\epsfig{file=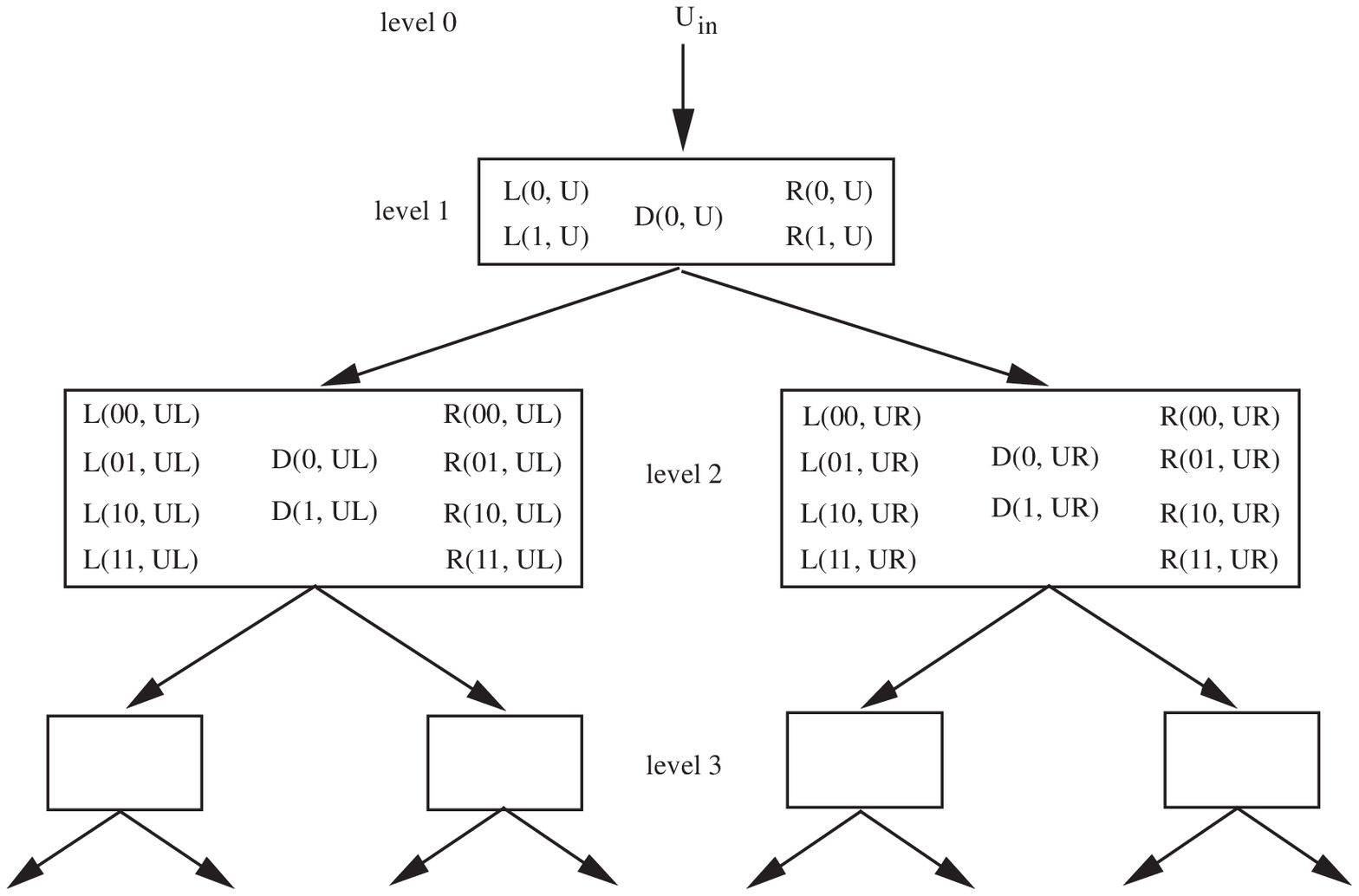}
			
			{Fig.1 A CSD binary tree.}
		\end{center}


Consider Fig.1. 
We start with an initial unitary matrix $U_{in}$ at horizontal level 0.
Without loss of generality, we can assume that 
the dimension of $U_{in}$ is $2^\nb$ for some $\nb\geq 1$.
(If initially $U_{in}$'s dimension is not a power of 2,
 we replace it by a direct sum
$U_{in}\oplus I_r$ whose dimension is a power of two.)
We apply the CSD method to $U_{in}$. This yields for level 1 a
D matrix $D(0, U)$, two unitary matrices 
$L(0, U)$ and $L(1, U)$ on the left side and two unitary
matrices $R(0, U)$ and $R(1, U)$ on the right side.
Then we apply the CSD method to 
each of the 4 matrices 
$L(0, U), L(1, U), R(0,U)$ and $R(1, U)$ that were produced
in the previous step. Then we apply the 
CSD method to each of the 16 $R$ and $L$ matrices that were produced
in the previous step. And so on. 
Each node $\caln$ has children $\caln_L$ and $\caln_R$ to
its left and right, respectively.
$\caln_R$ (ditto, $\caln_L$) has 
$D, R, L$ matrices produced by applying the CSD method
to the $R$ matrices (ditto, $L$ matrices) of its parent node $\caln$.
At level $\nb$, the
$L$'s and $R$'s are $1\times 1$ dimensional---i.e., 
just unit-modulus complex numbers. 
The nodes of the last level $\nb + 1$
don't have $R, L$ or $D$ matrices per se.
If $\caln_R$ (ditto, $\caln_L$) is a node of level $\nb + 1$, then it
stores the $R$ (ditto, $L$)
matrices of its parent node $\caln$.

Call a {\it central matrix} either (1) a single D matrix, or 
(2) a direct sum $D_1 \oplus D_2 \oplus \cdots  \oplus D_r$ of D matrices,
or (3) a diagonal unitary matrix. From Fig.1 it is clear that 
the initial matrix $U_{in}$ can be expressed as a product of 
central matrices, with each node of the tree providing
one of the central matrices in the product. Later on we will 
present techniques
for decomposing any central matrix into a SEO.

\section*{2. Preliminaries}
\mbox{}\indent	
In this section, we introduce some notation and some general mathematical
concepts that will be used in subsequent sections.

\subsection*{2(a) General Notation}
\mbox{}\indent	
We define
$Z_{a, b} = \{ a, a+1, \ldots, b\}$ for any integers $a$ and $b$.
$\delta(x,y)$ equals one if $x=y$ and zero otherwise.

We will use the symbol $\nb$ for the number ($\geq 1$) of bits and 
$\ns = 2^\nb$ for the number of states with $\nb$ bits. 
Let $Bool = \{0,1\}$. We will use lower case Latin letters
$a,b,c\ldots \in Bool$ to represent bit values and
lower case Greek letters $\alpha, \beta, \gamma, \ldots \in \bitz $
to represent bit positions. A vector such as 
$\va = a_{\nb-1} \ldots a_2 a_1 a_0$ will represent 
a string of bit values, $a_\mu$ being the value of the $\mu$'th bit
for $\mu \in \bitz$. A bit string $\va$ has
a decimal representation 
$d(\va) = \sum^{\nb-1}_{\mu=0} 2^\mu a_\mu$.
For $\beta\in \bitz$, we will use 
$\vec{u}(\beta)$ to denote the $\beta$'th standard unit vector---i.e., 
the vector with bit value of 1 at bit position $\beta$ and
bit value of zero at all other bit positions.

The set $Bool^\nb$ can be ordered in the standard way,
the {\it dictionary ordering}. For example, 
the dictionary ordering of $Bool^3$ is

\beq
(000, 001, 010, 011, 100, 101, 110, 111)
\;.
\eqlabel{2a.1}\eeq
Other useful orderings of $Bool^\nb$
are the so called {\it Gray codes}\enote{Gray}, named, not after the color,
but after an actual person named Gray.
In Gray codes, the next bit string may only differ from 
its predecessor in the value of a single bit.
Clearly, this condition does not specify a unique ordering
of $Bool^\nb$, so there is more than one Gray code of $Bool^\nb$.
Henceforth, we will refer to a Gray code as a {\it lazy ordering},
because, as we step from any $\vb$ to the next,
we act ``lazily", flipping only one bit instead of many.
An example of a lazy ordering of $Bool^3$ is

\beq
(000, 100, 110, 010, 011, 111, 101, 001)
\;.
\eqlabel{2a.2}\eeq

We define the single-qubit states $\ket{0}$ and $\ket{1}$ by
 
\beq
\ket{0} =
\left[
\begin{array}{c}
1 \\ 0 
\end{array}
\right]
\;\;,\;\;
\ket{1} =
\left[
\begin{array}{c}
0 \\ 1 
\end{array}
\right]
\;.
\eqlabel{2a.3}\eeq
If $\va \in Bool^{\nb}$, we define the 
$\nb$-qubit state $\ket{\va}$ as the following tensor product

\beq
\ket{\va} = \ket{a_{\nb -1}} 
\otimes \ldots \ket{a_1} \otimes \ket{a_0}
\;.
\eqlabel{2a.4}\eeq
For example, 

\beq
\ket{01} = 
\left[
\begin{array}{c}
1 \\ 0 
\end{array}
\right]
\otimes
\left[
\begin{array}{c}
0 \\ 1 
\end{array}
\right]
=
\left[
\begin{array}{c}
0 \\ 1 \\ 0 \\0
\end{array}
\right]
\;.
\eqlabel{2a.5}\eeq

$I_r$ will represent the $r$ dimensional unit matrix.
Suppose $\beta\in \bitz$ and $M$ is any $2\times 2$ matrix. We define
$M(\beta)$ by
\beq
M(\beta) = 
I_2 \otimes
\cdots \otimes 
I_2 \otimes 
M \otimes 
I_2 \otimes
\cdots \otimes 
I_2
\;,
\eqlabel{2a.6}\eeq
where the matrix $M$ on the right side is located
at bit position $\beta$ in the tensor product 
of $\nb$ $2\times 2$ matrices.
The numbers that label bit positions in the 
tensor product increase from
right to left ($\leftarrow$), and the rightmost bit is taken
to be at position 0.

For any pair of same-sized square matrices $A$ and $B$ such that
$\det(B)\neq 0$, one can define $B^A = \exp(A \ln B)$ using 
the Taylor series of $\ln(\cdot)$ about the identity matrix and
the Taylor series of $\exp(\cdot)$ about zero. This gives
a definition for objects such as $M_1(\beta_1)^{M_2(\beta_2)}$,
where $M_1$ and $M_2$ are $2\times 2$ matrices, $\det(M_1)\neq 0$ and
$\beta_1, \beta_2 \in \bitz$.

For any two same-sized square matrices $A$ and $B$,
 we define the o-dot product $\odot$  by
 $A \odot B = A B A^{\dagger}$, where 
$A^{\dagger}$ is the Hermitian conjugate of $A$.

$\vec{\sigma} = (\sx, \sy , \sz)$ will 
represent the vector of Pauli matrices, where

\beq
\sx=
\left(
\begin{array}{cc}
0&1\\
1&0
\end{array}
\right)
\;,
\;\;
\sy=
\left(
\begin{array}{cc}
0&-i\\
i&0
\end{array}
\right)
\;,
\;\;
\sz=
\left(
\begin{array}{cc}
1&0\\
0&-1
\end{array}
\right)
\;.
\eqlabel{2a.7}\eeq
A {\it qubit rotation} is defined as any matrix 
of the form
$\exp[i\vec{\theta}\cdot\vec{\sigma}(\beta)]$, 
where $\beta \in \bitz$ and $\vec{\theta}$ is a real 
3-dimensional vector.
 
\subsection*{2(b) Projection Operators}
\mbox{}\indent	
Consider a single qubit first.

The number operator $n$ of the qubit is defined by

\beq
n = 
\left[
\begin{array}{cc}
0 & 0 \\
0 & 1
\end{array}
\right]
=
\frac{ 1 - \sz}{2}
\;.
\eqlabel{2b.1}\eeq
Note that 

\beq
n \ket{0} = 0\ket{0} = 0
\;\;,\;\;
n \ket{1} = 1\ket{1}
\;.
\eqlabel{2b.2}\eeq
We will often use $\nbar$ as shorthand for 

\beq
\nbar =
1-n = 
\left[
\begin{array}{cc}
1 & 0 \\
0 & 0
\end{array}
\right]
=
\frac{ 1 + \sz}{2}
\;.
\eqlabel{2b.3}\eeq
Define $P_0$ and $P_1$ by

\beq
P_0 = \nbar =
\left[
\begin{array}{cc}
1 & 0 \\
0 & 0
\end{array}
\right]
\;\;,\;\;
P_1 = n =
\left[
\begin{array}{cc}
0 & 0 \\
0 & 1
\end{array}
\right]
\;.
\eqlabel{2b.4}\eeq
$P_0$ and $P_1$ are orthogonal projectors and they add to one:

\beq
P_a P_b = \delta(a, b) P_b
\;\;\;\;\; {\rm for} \;\; a,b\in Bool
\;,
\eqlabel{2b.5}\eeq

\beq
P_0 +  P_1 = I_2
\;.
\eqlabel{2b.6}\eeq

Now consider $\nb$ bits instead of just one. 

For 
$\beta\in \bitz$, we define $P_0(\beta)$, $P_1(\beta)$, $n(\beta)$
and $\nbar(\beta)$ according to Eq.(2a.6).\enote{Havel}

For $\va \in Bool^\nb$, let

\beq
P_{\va} = P_{a_{\nb-1}} \otimes \cdots
\otimes P_{a_2} \otimes P_{a_1} \otimes P_{a_0}
\;.
\eqlabel{2b.7}\eeq
For example,
with 2 bits we have

\beq
P_{00} = P_0 \otimes P_0 = diag(1, 0, 0, 0)
\;,
\eqlabel{2b.8a}\eeq

\beq
P_{01} = P_0 \otimes P_1 = diag(0, 1, 0, 0)
\;,
\eqlabel{2b.8b}\eeq

\beq
P_{10} = P_1 \otimes P_0 = diag(0, 0, 1, 0)
\;,
\eqlabel{2b.8c}\eeq

\beq
P_{11} = P_1 \otimes P_1 = diag(0, 0, 0, 1)
\;.
\eqlabel{2b.8d}\eeq
Note that 

\beq
P_\va P_\vb = \delta(\va, \vb) P_\vb
\;\;\;\;\; {\rm for} \;\; \va,\vb\in Bool^\nb
\;,
\eqlabel{2b.9}\eeq

\beq
\sum_{\va\in Bool^\nb } 
P_{\va} =
I_2 \otimes I_2 \otimes \cdots \otimes I_2 = I_{2^\nb}
\;.
\eqlabel{2b.10}\eeq

For $r\geq 1$, suppose $P_1, P_2, \ldots P_r$ are orthogonal
projection operators (i.e., $P_i P_j = \delta(i, j) P_j$ ), and
$\alpha_1, \alpha_2 \ldots \alpha_r$ are complex numbers. Then it is easy
to show by Taylor expansion that 

\beq
\exp ( \sum_{i=1}^r \alpha_i P_i)
=
\sum_{i=1}^r \exp(\alpha_i) P_i +
( 1 - \sum_{i=1}^r P_i )
\;.
\eqlabel{2b.11}\eeq
In other words, one can ``pull out" the summation sign
 from the argument of the exponential, but
only if one adds a compensating term $1 - \sum_i P_i$ so that 
both sides of the equation agree when all the $\alpha_i$'s are zero. 

For any pair of $2\times 2$ matrices $\tau_0, \tau_1$, define

\beq
\tau_\vb = \tau_{b_{\nb-1}}\otimes \ldots \otimes \tau_{b_1} \otimes \tau_{b_0} 
\;,
\eqlabel{2b.12}\eeq
where $\vb\in Bool^\nb$. Now let

\beq
{\cal B}(\tau_1, \tau_0) =
\{ \tau_\vb  |
\vb \in Bool^\nb \}
\;.
\eqlabel{2b.13}\eeq
The set of $\ns \times \ns$ diagonal complex matrices is a vector
space and ${\cal B}(P_1, P_0) = {\cal B}(n, \nbar)$ is a basis for it.
${\cal B}(\sz, I_2)$ and ${\cal B}(n, I_2)$ are also bases for it.

\subsection*{2(c) Sylvester-Hadamard Matrices}
\mbox{}\indent	
The $\nb$-bit Sylvester-Hadamard matrix\enote{Had} $H_\nb$ is 
defined by:

\beq
H_1 = 
\begin{tabular}{r|rr}
         & {\tiny 0} & {\tiny 1} \\
\hline
{\tiny 0}& 1&  1\\
{\tiny 1}& 1& -1\\
\end{tabular}
\;,
\eqlabel{2c.1a}\eeq

\beq
H_2 = 
\begin{tabular}{r|rrrr}
         & {\tiny 00} & {\tiny 01} & {\tiny 10} & {\tiny 11}\\
\hline
{\tiny 00}&  1&  1&  1&  1\\
{\tiny 01}&  1& -1&  1& -1\\
{\tiny 10}&  1&  1& -1& -1\\
{\tiny 11}&  1& -1& -1&  1\\
\end{tabular}
\;,
\eqlabel{2c.1b}\eeq

\beq
H_{r+1} = H_1 \otimes H_r
\;,
\eqlabel{2c.1c}\eeq
for any integer $r\geq 1$. We will often use a plain $H$
to represent $H_1$. In Eqs.(2c.1), 
 we have labelled the rows and columns with binary numbers
 in increasing dictionary order.
From Eqs.(2c.1), one can show that
the entry of $H_\nb$ at row $\va \in Bool^\nb$ and column $\vb\in Bool^\nb$ is given by

\beq
(H_\nb)_{\va, \vb} = (-1)^{\va\cdot \vb}
\;,
\eqlabel{2c.2}\eeq
where 
$\va\cdot \vb = \sum^{\nb-1}_{\mu=0} a_\mu b_\mu$. It is easy to 
check that

\beq
H_\nb^T = H_\nb
\;,
\eqlabel{2c.3}\eeq

\beq
H_\nb^2 = \ns I_\ns
\;.
\eqlabel{2c.4}\eeq
In other words, $H_\nb$ is a symmetric matrix, and 
the inverse of $H_\nb$ equals
$H_\nb$ divided by however many rows it has.

If $H(\beta)$ for $\beta\in \bitz$ is defined 
according to Eq.(2a.6), then Eqs.(2c.1) imply that

\beq
H_\nb = H(\nb-1)\ldots H(2) H(1) H(0)
\;.
\eqlabel{2c.5}\eeq
The $H(\beta)$'s on the right side of the last equation commute so also

\beq
H_\nb = H(0)H(1)H(2)\ldots H(\nb-1)
\;.
\eqlabel{2c.6}\eeq

\subsection*{2(d) Discrete Fourier Transform}
\mbox{}\indent	
The $\nb$ bit Discrete Fourier Transform (DFT) matrix $F_\nb$
is defined by

\beq
(F_\nb)_{a,b} = \frac{1}{\sqrt{\ns}} \omega^{ab}
\;,
\eqlabel{2d.1}\eeq
where $a, b \in Z_{0, \ns-1}$ and

\beq
\omega = e^{i \frac{2\pi}{\ns} }
\;.
\eqlabel{2d.2}\eeq
Note that $F_\nb$ is a symmetric matrix. It is easy
to show that it is also a unitary matrix. If
$\tilde{v}$ and $v$ are complex $\ns$ dimensional vectors such that
$\tilde{v} = F_\nb v$, then we call $\tilde{v}$ the DFT of $v$.

Calculating $\tilde{v}$ the naive way, by multiplying
$v$ by $F_\nb$, would take Order($\ns^2$) 
classical elementary operations (complex multiplications
mostly). Instead, it is possible to calculate $\tilde{v}$ from
$v$ in Order($\ns \ln \ns$) classical elementary operations
using the well known\enote{Knuth} Fast Fourier Transform (FFT) algorithm.
Ref.\enote{Copper} was the first to
 express this algorithm as a product of 
matrices each of which acts on at most 2 bits at a time. This way of 
expressing it is often called the 
``quantum FFT algorithm" because it is ideal for quantum computation.
Ref.\enote{Copper} showed (See Appendix B) that  

\beq
F_\nb = \frac{1}{\sqrt{\ns}}
H(\nb - 1) \ldots \Delta(2) H(2) \Delta(1) H(1) \Delta(0) H(0) P_{BR}
\;,
\eqlabel{2d.3}\eeq
where $H(\alpha)$ is the 1-bit Hadamard matrix operating on
bit $\alpha\in \bitz$, $P_{BR}$ is the bit reversal 
matrix for $\nb$ bits, and 

\beq
\Delta(\beta) = 
\Delta(\beta + 1, \beta) 
\Delta(\beta + 2, \beta) \ldots
\Delta(\nb - 1, \beta) 
\;,
\eqlabel{2d.4}\eeq
where

\beq
\Delta(\alpha, \beta) = 
\exp[
i \phi_{|\alpha - \beta| + 1}
n(\alpha) n(\beta) ]
\;,
\eqlabel{2d.5}\eeq

\beq
\phi_\gamma = \frac{2 \pi}{2^\gamma}
\;.
\eqlabel{2d.6}\eeq

Thus, $\Delta(\alpha, \beta)$ is a diagonal
matrix whose diagonal entries are either 1 or a phase factor.
For example, for $\nb = 3$, 

\beq
n(0)n(2) \ket{a_2, a_1, a_0 } =
\left \{
\begin{array}{l}
1\ket{a_2, a_1, a_0} \;\;\;{\rm if} \;\; a_2 = a_0 = 1 \\
0 \;\;\;{\rm otherwise}
\end{array}
\right.
\;,
\eqlabel{2d.7}\eeq
so 

\beq
\Delta(0, 2) = e^{i \phi_3 n(0)n(2)} =
diag(
\stackrel{000}{1}, 
\stackrel{001}{1}, 
\stackrel{010}{1}, 
\stackrel{011}{1}, 
\stackrel{100}{1}, 
\stackrel{101}{e^{i\phi_3}}, 
\stackrel{110}{1}, 
\stackrel{111}{e^{i\phi_3}}
) 
\;.
\eqlabel{2d.8}\eeq

For $\nb = 3$, reversing the bits of the numbers contained in $Z_{0,7}$
exchanges $1 = d(001)$ with $4 = d(100)$ and
$3 = d(011)$ with $6 = d(110)$, and it leaves all other 
numbers in $Z_{0, 7}$ the same. Thus, for $\nb = 3$, $P_{BR}$ is the 
$8\times 8$ permutation matrix which corresponds to the
following product of transpositions: (1,4)(3,6).

Note that $F_\nb$, $H(\alpha)$, $\Delta(\alpha)$ and
$P_{BR}$ are all symmetric matrices. Hence, taking the 
transpose of both sides of Eq.(2d.3), one gets

\beq
F_\nb = \frac{1}{\sqrt{\ns}}
    P_{BR} H(0) \Delta(0) H(1) \Delta(1) H(2) \Delta(2) \ldots H(\nb - 1)
\;.
\eqlabel{2d.9}\eeq
Both the last equation and Eq.(2d.3) are called the quantum FFT algorithm.

		\begin{center}
			\epsfig{file=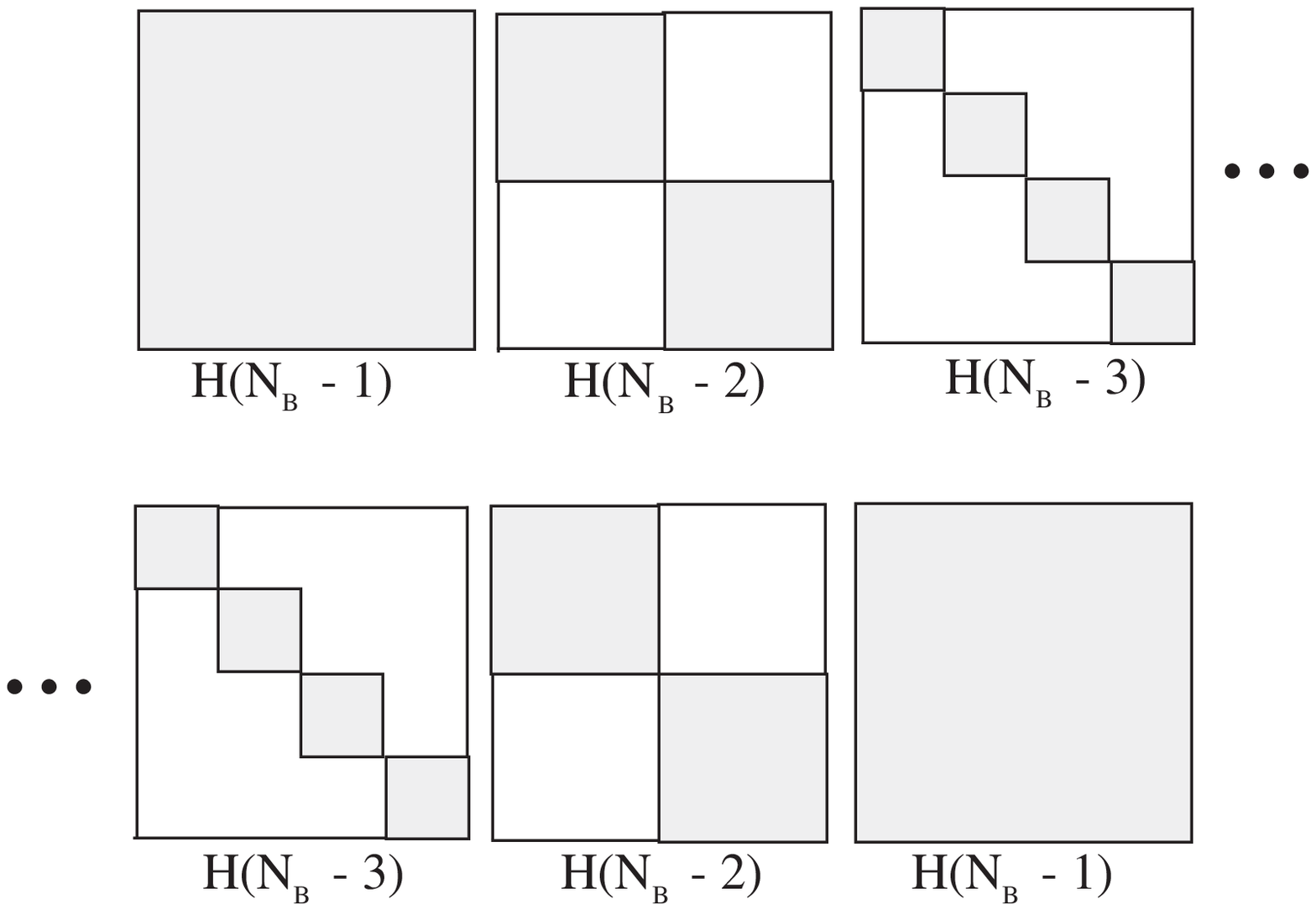}
			
			{Fig.2 Pictorial representation of
			quantum FFT algorithm. For simplicity, we
			only show the Hadamard matrices.
			White (unshaded) regions inside a matrix
			represent zero entries.}
		\end{center}


Fig.2 is a pictorial representation of Eqs.(2d.3) and (2d.9).
For any pair of $2\times 2$ complex matrices $M, M'$
and for $\beta, \beta'\in \bitz$, we will say that
$M(\beta)$ is ``more diagonal" than $M'(\beta)$ if
$\beta < \beta'$. $M(0)$ is a direct sum of $2\times 2$ matrices.
$M(1)$ is a direct sum of $4 \times 4$ matrices so it is 
less diagonal than $M(0)$, etc. The matrices $\Delta(\alpha)$
that occur in Eqs.(2d.3) and (2d.9) are truly diagonal but
the matrices $H(\alpha)$ aren't. In Eq.(2d.3) (ditto, Eq.(2d.9))
the $H(\alpha)$ matrices become less diagonal (ditto, more diagonal)
as one goes from right to left.

 The fact that the non-diagonal
matrices $H(\alpha)$ are real and the diagonal matrices 
$\Delta(\alpha)$ are complex is what gives the FFT
algorithm its speed advantage in classical computation.
Of course, segregating real and complex operations has no 
advantage in quantum computation, where complex numbers are
very natural. The charm of the FFT algorithm in quantum
computation is that it can be expressed as a short 
SEO, namely Eqs.(2d.3) and (2d.9).

\subsection*{2(e) Permutations}
\mbox{}\indent	
Subsequent sections will use the following very 
basic facts about permutations.
For more details, see, for example, Ref.\enote{Perm}.

A {\it permutation} is a 1-1 onto map from a finite set $X$ onto itself.
The set of permutations on set $X$ is a group if group multiplication
is taken to be function composition. $S_n$, the {\it symmetric group
in $n$ letters}, is defined as the group of all permutations
on any set $X$ with $n$ elements.
If $X= Z_{1,n}$, then a permutation $G$ which maps $i\in X$ to $a_i\in X$
(where $i\neq j$ implies $a_i\neq a_j$) can be represented by a 
matrix with entries

\beq
(G)_{j, i} = \delta(a_i, j)
\;,
\eqlabel{2e.1}\eeq
for all $i,j \in X$. Note that all entries in any given row or column 
equal zero except for one entry which equals one.
Hence, the rows of $G$ are orthonormal and $G$
is an orthogonal matrix ($G^T G = G G^T = 1$).
 An alternative notation for $G$ is 

\beq
G = 
\left (
\begin{array}{ccccc}
1   & 2   & 3   & \cdots & n   \\
a_1 & a_2 & a_3 & \cdots & a_n
\end{array}
\right )
\;.
\eqlabel{2e.2}\eeq
The product of two symbols of the type shown in Eq.(2e.2) 
is defined by function composition. For example,

\beq
\left (
\begin{array}{ccc}
a_1 & a_2 & a_3 \\
b_1 & b_2 & b_3 
\end{array}
\right )
\left (
\begin{array}{ccc}
  1 &   2 &   3 \\
a_1 & a_2 & a_3 
\end{array}
\right )
=
\left (
\begin{array}{ccc}
  1 &   2 &   3 \\
b_1 & b_2 & b_3 
\end{array}
\right )
\;.
\eqlabel{2e.3}\eeq
Note how we have applied the 
permutations on the left side of the equation
from right to left ($\leftarrow$). (Careful: Some authors apply them in
the opposite direction ($\rightarrow$)). A cycle is a special type of
permutation. If $G\in S_n$ maps 
$a_1\rarrow a_2$, $a_2\rarrow a_3$, \ldots, $a_{r-1}\rarrow a_r$,
$a_r\rarrow a_1$, where $i\neq j$ implies $a_i\neq a_j$ and $r\leq n$,
then we call $G$ a {\it cycle}. $G$ may be
 represented as in Eqs.(2e.1) and (2e.2).
Another way to represent it is by

\beq
G = ( a_1, a_2, a_3, \ldots, a_r)
\;.
\eqlabel{2e.4}\eeq
(Careful: some people write $( a_r, \ldots, a_3, a_2, a_1)$ instead.)
We say that the cycle of Eq.(2e.4) has {\it length} $r$. 
Cycles of length 1 are
just the identity map. A cycle of length 2 is 
called a {\it transposition}.
The product of two cycles need not be another cycle. For example,

\beq
(2,1,5) (1,4,5,6) =
\left ( 
\begin{array}{cccccc}
1 & 2 & 3 & 4 & 5 & 6\\
4 & 1 & 3 & 2 & 6 & 5 
\end{array}
\right )
\;
\eqlabel{2e.5}\eeq
cannot be expressed as a single cycle. Any permutation
can be written as a product of cycles. For example,

\beq
\left ( 
\begin{array}{cccccc}
1 & 2 & 3 & 4 & 5 & 6\\
4 & 1 & 3 & 2 & 6 & 5 
\end{array}
\right )
=
(5,6) (1,4,2)
\;.
\eqlabel{2e.6}\eeq
The cycles on the right side of Eq.(2e.6) are {\it disjoint}; i.e., they
have no elements in common. Disjoint cycles commute. Any cycle  
can be expressed as a product of transpositions 
(assuming a group with $\geq 2$ elements),
 by using identities such as:

\beq
(a_1, a_2, \ldots, a_n) = 
(a_1, a_2) (a_2, a_3) \cdots (a_{n-1}, a_n)
\;,
\eqlabel{2e.7}\eeq

\beq
(a_1, a_2, \ldots, a_n) = 
(a_1, a_n) \cdots (a_1, a_3)(a_1, a_2)
\;.
\eqlabel{2e.8}\eeq
Another useful identity is

\beq
(a, b) = (a, p)(p, b)(a, p)
\;.
\eqlabel{2e.9}\eeq
This last identity can be applied repeatedly. For example, applied 
twice, it gives

\beq
(a, b) = (a, p_1)(p_1, b)(a, p_1) = (a, p_1) (p_1, p_2) (p_2, b) (p_1, p_2) (a, p_1)
\;.
\eqlabel{2e.10}\eeq
Since any permutation equals a product of cycles, and each 
of those cycles can be expressed as a product of transpositions,
all permutations can be 
expressed as a product of transpositions (assuming 
a group with $\geq 2$ elements). 
The decomposition of a 
permutation into transpositions is not unique. However,
the number of transpositions whose product equals a given
permutation is always either even or odd. An 
{\it even} (ditto, {\it odd}) {\it permutation} is
defined as one which equals an even (ditto, odd) 
number of transpositions.

\section*{3. State Permutations that Act on Two Bits}
\mbox{}\indent	
The goal of this paper is to reduce any unitary matrix into 
a product of qubit rotations and 
controlled-nots (c-nots). A qubit rotation
acts on 
a single qubit at a time. This section will discuss 
gates such as c-nots that are state permutations that  
act on two bits at a time.

\subsection*{3(a) $\nb = 2$}
\mbox{}\indent	
Consider first the case when there are only 2 bits. Then there
are four possible states--00, 01, 10, 11. With these 4 states,
one can build 6 distinct transpositions:

\beq
(00, 01) = 
\left [
\begin{array}{cc}
\sx 	&   \\
 		& I_2 
\end{array}
\right ]
=
P_0 \otimes \sx
+
P_1 \otimes I_2
=
\cnotno{1}{0}
\;,
\eqlabel{3a.1a}\eeq

\beq
(00, 10) = 
\left [
\begin{array}{cc}
P_1 	& P_0  \\
P_0 	& P_1 
\end{array}
\right ]
=
I_2 \otimes P_1
+
\sx \otimes P_0
=
\cnotno{0}{1}
\;,
\eqlabel{3a.1b}\eeq

\beq
(00, 11) = 
\left [
\begin{array}{ccc}
  &     & 1  \\
  & I_2 &    \\
 1&     &
\end{array}
\right ]
\;,
\eqlabel{3a.1c}\eeq

\beq
(01, 10) = 
\left [
\begin{array}{ccc}
 1 &     &   \\
  & \sx &    \\
 &     & 1
\end{array}
\right ]
\;,
\eqlabel{3a.1d}\eeq

\beq
(01, 11) = 
\left [
\begin{array}{cc}
P_0 	& P_1  \\
P_1 	& P_0 
\end{array}
\right ]
=
I_2 \otimes P_0
+
\sx \otimes P_1
=
\cnotyes{0}{1}
\;,
\eqlabel{3a.1e}\eeq

\beq
(10, 11) = 
\left [
\begin{array}{cc}
I_2 	&   \\
 	    & \sx 
\end{array}
\right ]
=
 P_0 \otimes I_2 
+
 P_1 \otimes \sx
=
\cnotyes{1}{0}
\;,
\eqlabel{3a.1f}\eeq
where matrix entries left blank should be interpreted as zero.
The rows and columns of the above matrices are labelled by 
binary numbers in increasing dictionary order (as in Eq.(2c.1b) for $H_2$).
Note that the 4 transpositions Eqs.(3a.1)(a,b,e,f) change 
only one bit value. We will call them {\it controlled nots (c-nots)}. 
The other 2 transpositions 
Eqs.(3a.1)(c,d) change both bit values. We will call 
$(00, 11)$ the {\it Twin-to-twin-er} and $(01, 10)$ 
the {\it Exchanger}.
Expressions such as $\cnotyes{\alpha}{\beta}$ 
where $\alpha \neq \beta$
are a special case of $M_1(\beta_1)^{M_2(\beta_2)}$,
which was defined in Section 2a.
$\cnotyes{\alpha}{\beta}$ equals 
$\sx(\beta)$ when it acts on a state for which $n(\alpha)=1$,
whereas it equals 1 if $n(\alpha)=0$. $\alpha$ is called
the {\it control bit}  and $\beta$ the {\it flipper bit}.

Exchanger\enote{Fey} has four possible representations as
a product of c-nots:

\beq
(01, 10) = (01, 00) (00, 10) (01, 00) = 
\cnotno{1}{0}\cnotno{0}{1} \cnotno{1}{0}
\;,
\eqlabel{3a.2a}\eeq

\beq
(01, 10) = (10, 11) (11, 01) (10, 11) = 
\cnotyes{1}{0}\cnotyes{0}{1} \cnotyes{1}{0}
\;,
\eqlabel{3a.2b}\eeq

\beq
(01, 10) = (10, 00) (00, 01) (10, 00) = 
\cnotno{0}{1}\cnotno{1}{0} \cnotno{0}{1}
\;,
\eqlabel{3a.2c}\eeq

\beq
(01, 10) = (01, 11) (11, 10) (01, 11) = 
\cnotyes{0}{1}\cnotyes{1}{0} \cnotyes{0}{1}
\;.
\eqlabel{3a.2d}\eeq
Note that one can go 
from Eq.(3a.2a) to (3a.2b) 
by exchanging $n$ and $\nbar$;
from Eq.(3a.2a) to (3a.2c)
by exchanging bit positions 0 and 1;
from Eq.(3a.2a) to (3a.2d) by doing both,
exchanging $n$ and $\nbar$ and
exchanging bit positions 0 and 1. We will often represent Exchanger
by $E(0,1)$. It is easy to show that

\beq
E^T(0,1) = E(0,1) = E^{-1}(0, 1)
\;,
\eqlabel{3a.3a}\eeq

\beq
E(0,1) = E(1,0)
\;,
\eqlabel{3a.3b}\eeq

\beq
E^2(0,1) = 1
\;.
\eqlabel{3a.3c}\eeq
Furthermore, if $X$ and $Y$ are two arbitrary $2\times2$ matrices, 
then, by using the matrix representation Eq.(3a.1d) of Exchanger,
one can show that 

\beq
E(1,0)\odot (X\otimes Y) = Y\otimes X
\;.
\eqlabel{3a.4}\eeq
Thus, Exchanger exchanges the position of matrices $X$ and $Y$
in the tensor product.

Twin-to-twin-er also has 4 possible representations as a product of
c-nots. One is

\beq
(00, 11) = (00, 01) (01, 11) (00, 01) = 
\cnotno{1}{0}\cnotyes{0}{1} \cnotno{1}{0}
\;.
\eqlabel{3a.5}\eeq
As with Exchanger, the other 3 representations are obtained by
exchanging: (1) $n$ and $\nbar$, (2) bit positions 0 and 1, (3) both.

		\begin{center}
			\epsfig{file=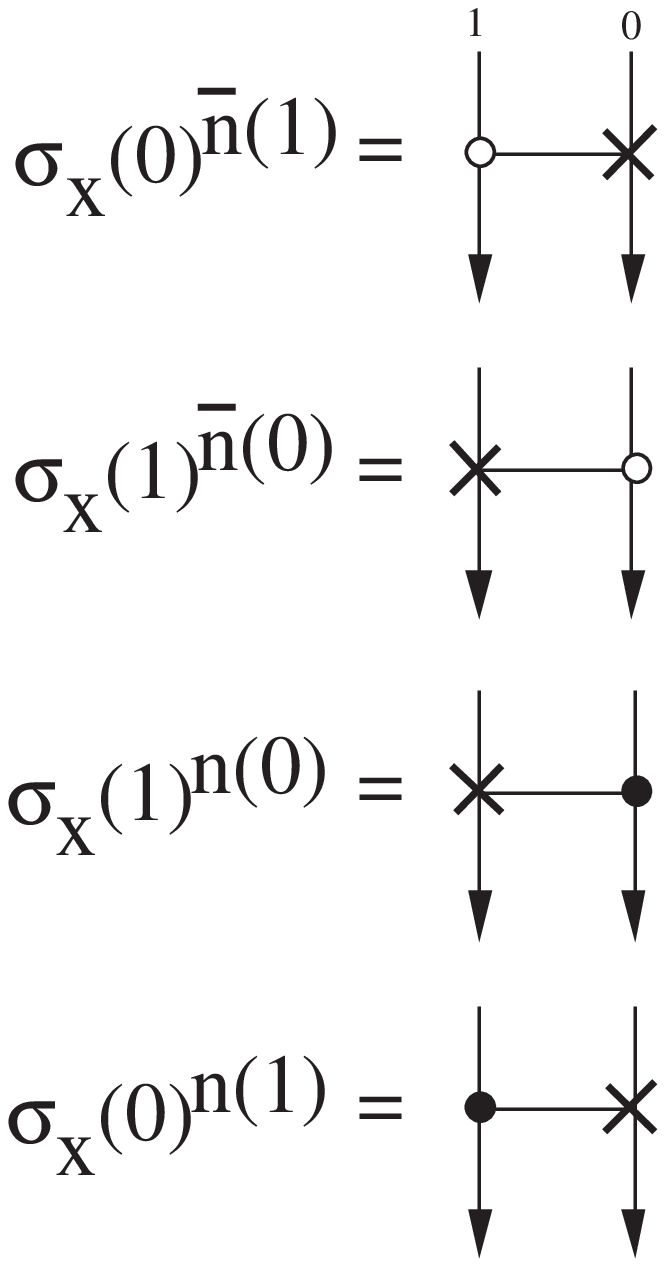}
			
			{Fig.3 Circuit symbols for the 4 different types of c-nots.}
		\end{center}

		\begin{center}
			\epsfig{file=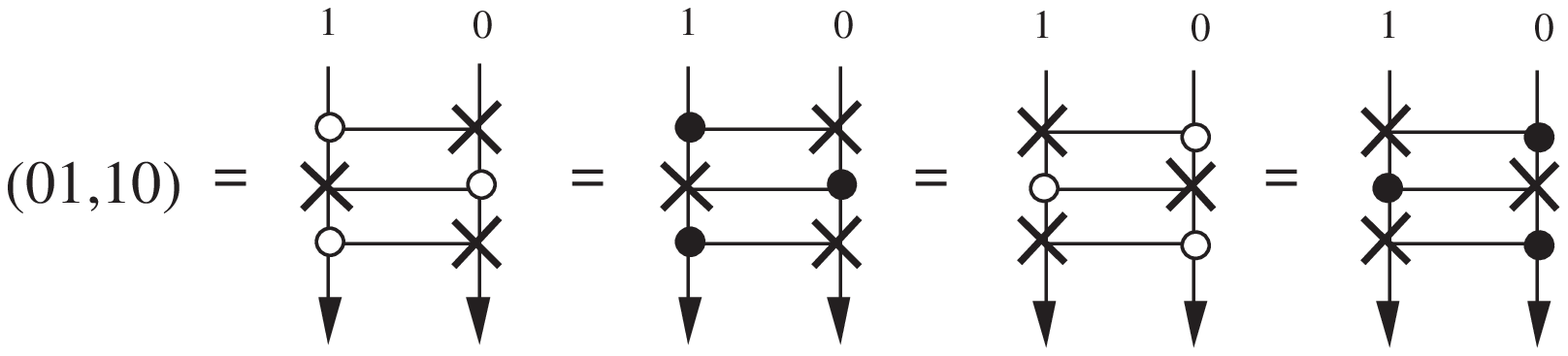}
			
			{Fig.4 Four equivalent circuit diagrams for Exchanger.}
		\end{center}

		\begin{center}
			\epsfig{file=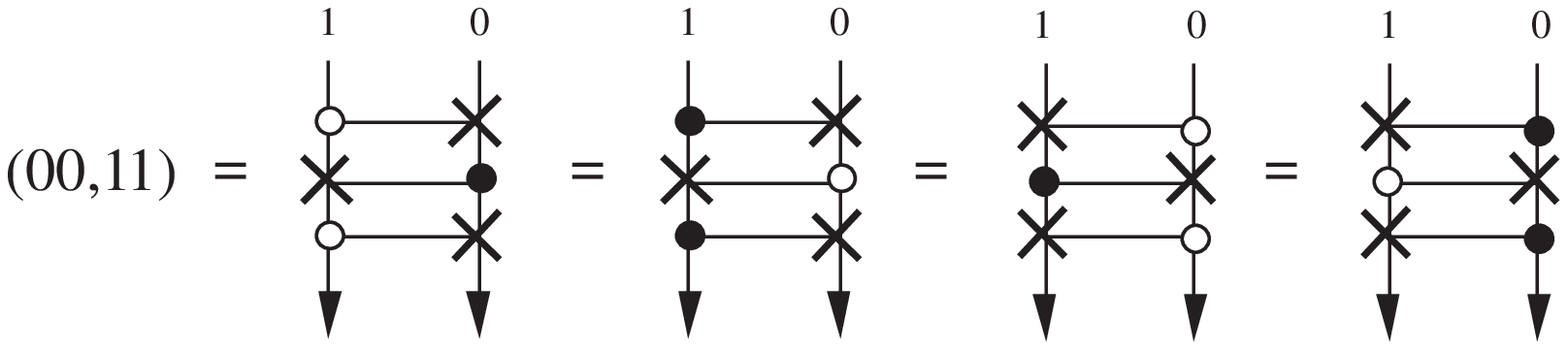}
			
			{Fig.5 Four equivalent circuit diagrams for Twin-to-twin-er.}
		\end{center}


Figures 3, 4 and 5 give a diagrammatic representation of the 6 
possible transpositions of states for $\nb=2$.

\subsection*{3(b) Any $\nb \geq 2$}
\mbox{}\indent
Suppose $a_1, b_1, a_2, b_2 \in Bool$ and
$\alpha, \beta \in \bitz$ such that $\alpha\neq \beta$. We define 

\beq
(a_1 b_1, a_2 b_2)_{\alpha, \beta}
=
\prod_{ (\Lambda, \Lambda', \Lambda'')\in Bool^{\nb -2} }
(\Lambda a_1 \Lambda' b_1 \Lambda'', \Lambda a_2 \Lambda' b_2 \Lambda'')
\;,
\eqlabel{3b.1}\eeq
where on the right side, $a_1, a_2$ are located at bit position $\alpha$,
and $b_1, b_2$ are located at bit position $\beta$. 
(Note that the transpositions
on the right side of Eq.(3b.1) are disjoint so they commute.)
For example, for $\nb=3$, 

\beq
\cnotyes{1}{0} =
(10, 11)_{1,0} =
\prod_{a\in Bool}
(a10, a11) =
(010, 011) (110, 111)
\;.
\eqlabel{3b.2}\eeq
Clearly, any permutation of states with $\nb$ bits
that acts on only 2 bits (i.e., Exchanger, 
Twin-to-twin-er, and all c-nots)
can be represented by
$(a_1 b_1, a_2 b_2)_{\alpha, \beta}$.

For $\alpha, \beta \in \bitz$ such that $\alpha\neq\beta$, let $E(\alpha, \beta)$ represent
Exchanger:

\beq
E(\alpha, \beta) = (01, 10)_{\alpha, \beta}
\;.
\eqlabel{3b.3}\eeq
As in the $\nb=2$ case, $E(\alpha, \beta)$ can be expressed as 
a product of c-nots in 4 different ways. One way is

\beq
E(\alpha, \beta) =
\cnotyes{\beta}{\alpha}
\cnotyes{\alpha}{\beta}
\cnotyes{\beta}{\alpha}
\;.
\eqlabel{3b.4}\eeq
The other 3 ways are obtained by exchanging: (1) $n$ and $\nbar$,
(2) bit positions $\alpha$ and $\beta$, (3) both. Again as in the 
$\nb=2$ case,

\beq
E^T(\alpha, \beta) = E(\alpha, \beta) = E^{-1}(\alpha, \beta)
\;,
\eqlabel{3b.5a}\eeq

\beq
E(\alpha, \beta) = E(\beta, \alpha)
\;,
\eqlabel{3b.5b}\eeq

\beq
E^2(\alpha, \beta) = 1
\;.
\eqlabel{3b.5c}\eeq
Furthermore, if $X$ and $Y$ are 
two arbitrary  $2\times 2$ matrices and 
$\alpha, \beta\in \bitz$ such that $\alpha\neq \beta$, then 

\beq
E(\alpha, \beta)\odot [X(\alpha) Y(\beta)]
=
X(\beta) Y(\alpha)
\;.
\eqlabel{3b.6}\eeq

Equation (3b.6) is an extremely useful result. It says that
$E(\alpha, \beta)$ is a transposition of bit positions. 
Furthermore, the $E(\alpha, \beta)$ generate the 
group of $\nb !$ permutations of bit
positions. (Careful: this is not the same as the group of $(2^{\nb})!$
permutations of states with $\nb$ bits.)

		\begin{center}
			\epsfig{file=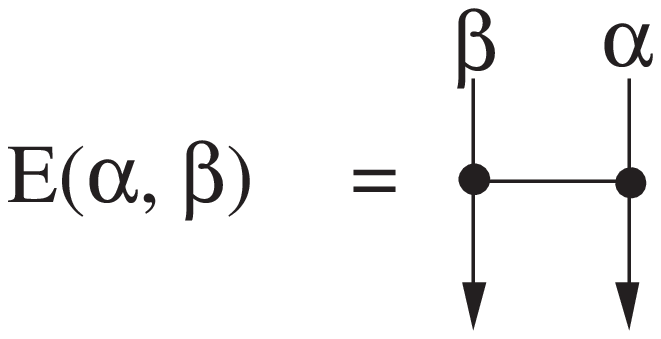}
			
			{Fig.6 Circuit symbol for Exchanger.}
		\end{center}

		\begin{center}
			\epsfig{file=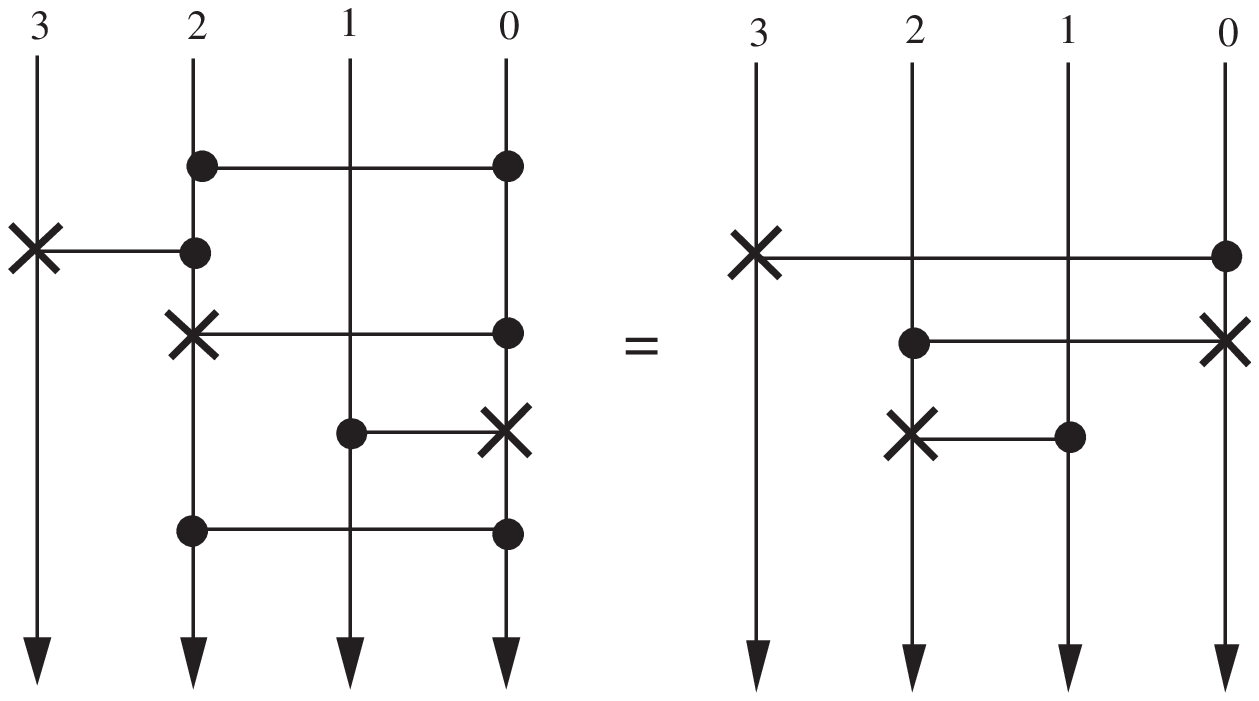}
			
			{Fig.7 Circuit diagram for Eq.(3b.7).}
		\end{center}


An example of 
how one can use Eq.(3b.6) is 

\beq
E(2, 0) \odot
[ \cnotyes{1}{0} \cnotyes{0}{2} \cnotyes{2}{3} ]
=
\cnotyes{1}{2} \cnotyes{2}{0} \cnotyes{0}{3}
\;.
\eqlabel{3b.7}\eeq
Figure 6 gives a convenient way of representing $E(\alpha, \beta)$
diagrammatically. Using this symbol, the example of Eq.(3b.7)
can be represented by Fig.7.

Certain experimental implementations of a quantum computer
might only allow
nearest-neighbor bit interactions.
Exchanger can be used to express
any elementary operation between 2 non-nearest-neighbor bits
as a SEO that contains only nearest-neighbor interactions.
For example, if $A(3,0)$ is an operator that acts on bits 3 and 0,
then

\beq
A(3,0) = [E(3,2) E(2,1)]\odot A(1,0)
\;.
\eqlabel{3b.8}\eeq

Of course, identities 
that are true for a general transposition are also true
for $E(\alpha, \beta)$. For example,

\beq
(2,0) = (2,1)(1,0)(2,1)
\;.
\eqlabel{3b.9}\eeq
Therefore,
\beq
E(2,0) = E(2,1)E(1,0)E(2,1)
\;.
\eqlabel{3b.10}\eeq

\section*{4. Decomposing Central Matrix into SEO}
\mbox{}\indent	
In Section 1(c), we gave only a partial description of our algorithm.
In this section, we complete that description by showing how to
decompose each of the 3 possible kinds of central matrices
into a SEO.

\subsection*{4(a)When Central Matrix is a Single D Matrix}
\mbox{}\indent	
D matrices are defined by Eqs.(1b.2). They can be expressed
in terms of projection operators as follows:

\beq
D = \sum_{\va \in Bool^{\nb-1}} \exp(i \phi_\va \sy )\otimes P_\va
\;,
\eqlabel{4a.1}\eeq
where the $\phi_\va$ are real numbers. Note that in Eq.(4a.1), $\va$ has
$\nb -1$ components instead of the full $\nb$. 
Using the identity Eq.(2b.11),
one gets

\beq
D =
\exp
\left (
i \sum_{\va\in Bool^{\nb - 1} }
\phi_{\va} 
\sy \otimes P_\va
\right )
\;.
\eqlabel{4a.2}\eeq
Now define new angles $\theta_\vb$ by

\beq
\phi_\va =
\sum_{\vb \in Bool^{\nb -1} }
(-1)^{\va\cdot\vb} \theta_\vb
\;.
\eqlabel{4a.3}\eeq
Suppose $\vec{\phi}$ (ditto, $\vec{\theta}$) is a column vector
whose components are the numbers $\phi_\va$ (ditto, $\theta_\va$)
arranged in order of increasing $\va$. Then Eq.(4a.3) is equivalent
to 

\beq
\vec{\phi} = H_{\nb -1} \vec{\theta}
\;.
\eqlabel{4a.4}\eeq
This is easily inverted to 

\beq
\vec{\theta} = 
\frac{1}{ 2^{\nb -1} }
H_{\nb - 1} \vec{\phi}
\;.
\eqlabel{4a.5}\eeq
Let $A_\vb$ for $\vb \in Bool^{\nb - 1}$ be defined by

\beq
A_\vb =
\exp
\left (
i \theta_\vb 
\sy
\otimes
\sum_{\va \in Bool^{\nb-1}}
(-1)^{\va\cdot\vb}
P_\va
\right )
\;.
\eqlabel{4a.6}\eeq
Then $D$ can be written as

\beq
D =
\prod_{ \vb \in Bool^{\nb -1} }
A_\vb
\;.
\eqlabel{4a.7}\eeq
Note that the $A_\vb$ operators on the right side commute
so the order in which they are multiplied is irrelevant.
Next we establish 2 useful identities:
If $\beta\in \bitz $ and $\vec{u}(\beta) \in Bool^{\nb }$
is the $\beta$'th standard unit vector, then 

\beq
\begin{array}{l}
\sum_{\va \in Bool^{\nb } }
(-1)^{\va \cdot \vec{u}(\beta) }
P_\va \\
= 
I_2 \otimes
\cdots\otimes
I_2\otimes
\left [\sum_{a_\beta \in Bool}
(-1)^{a_\beta}
P_{a_\beta}
\right ]
\otimes
I_2 \otimes
\cdots\otimes
I_2\\
=
P_0(\beta) - P_1(\beta)\\
=
\sz(\beta)
\end{array}
\;.
\eqlabel{4a.8}\eeq
If $\beta, \alpha \in \bitz $ and $\alpha \neq \beta$, then

\beq
\begin{array}{l}
\cnotyes{\alpha}{\beta}
\odot
\sy(\beta)\\
=
[ \sx(\beta) P_1(\alpha) + P_0(\alpha) ]
\odot
\sy(\beta)\\
=
\sy(\beta) 
[ - P_1(\alpha) + P_0(\alpha) ]\\
=
\sy(\beta) \sz(\alpha)
\end{array}
\;.
\eqlabel{4a.9}\eeq
Now we are ready to express $A_\vb$
in terms of elementary operators. For any $\vb \in Bool^{\nb -1} $, 
we can write

\beq
\vb = \sum_{j=0}^{r-1} \vec{u}(\beta_j)
\;,
\eqlabel{4a.10}\eeq
where 

\beq
\nb-2 \geq 
\beta_{r-1} 
> \cdots
> \beta_1 > \beta_0 \geq 0
\;.
\eqlabel{4a.11}\eeq
In other words, $\vb$ has bit value of 1 at bit positions 
$\beta_j$. At all other bit positions, $\vb$ has
bit value of 0. $r$ is the number of bits in $\vb$ whose
value is 1. When $\vb =0$, $r$ equals 0. 
By Eq.(4a.8)

\beq
A_\vb =
\exp
\left (
i \theta_\vb 
\sy(\nb -1)
\prod_{j=0}^{r-1}
\sz(\beta_j)
\right )
\;.
\eqlabel{4a.12}\eeq
By Eq.(4a.9), if $r\ge 1$,

\beq
\begin{array}{l}
[
\cnotyes{\beta_{r-1}}{\nb -1 }
\cdots
\cnotyes{\beta_{1}}{\nb -1 }
\cnotyes{\beta_{0}}{\nb -1 }
]
\odot
\sy (\nb -1)\\
=
\sy(\nb -1) 
\prod_{j=0}^{r-1}
\sz(\beta_j)\\
\end{array}
\;.
\eqlabel{4a.13}\eeq
Thus, for $r\ge 0$,

\beq
A_\vb
= 
[
\cnotyes{\beta_{r-1}}{\nb -1 }
\cdots
\cnotyes{\beta_{1}}{\nb -1 }
\cnotyes{\beta_{0}}{\nb -1 }
]
\odot
\exp[ i\theta_\vb \sy(\nb -1) ]
\;,
\eqlabel{4a.14}\eeq
where if $r=0$, the expression to the left of $\odot$ is 
defined to be 1.
There are other ways of decomposing $A_\vb$ into a SEO. For example,
using the above method, one can also show that

\beq
A_\vb
= 
[
\cnotyes{\beta_{r-1}}{\beta_{r-2} }
\cdots
\cnotyes{\beta_2}{\beta_1}
\cnotyes{\beta_1}{\beta_0}
\cnotyes{\beta_0}{\nb -1}
]
\odot
\exp[ i\theta_\vb \sy(\nb -1) ]
\;.
\eqlabel{4a.15}\eeq

In conclusion, we have shown how to decompose a D matrix into
a SEO. For example, suppose $\nb = 3$. Then

\beq
D = \sum_{a,b \in Bool}
\exp ( i \phi_{ab} \sy ) \otimes
P_a \otimes P_b
\;.
\eqlabel{4a.16}\eeq
Define $\vec{\theta}$ by 

\beq
\vec{\theta} = \frac{1}{4} H_2 \vec{\phi}
\;,
\eqlabel{4a.17}\eeq
and $\Gamma(\cdot)$ by 

\beq
\Gamma(\theta) = 
\exp( i \theta \sy )
\otimes I_2 \otimes I_2
\;.
\eqlabel{4a.18}\eeq
Then

\beq
D = A_{00} A_{01} A_{10} A_{11}
\;,
\eqlabel{4a.19}\eeq
where

\beq
A_{00} = \Gamma(\theta_{00})
\;,
\eqlabel{4a.20a}\eeq

\beq
A_{01} = 
\cnotyes{0}{2}\odot
\Gamma(\theta_{01})
\;,
\eqlabel{4a.20b}\eeq

\beq
A_{10} = 
\cnotyes{1}{2}\odot
\Gamma(\theta_{10})
\;,
\eqlabel{4a.20c}\eeq

\beq
A_{11} = 
[\cnotyes{1}{2}
\cnotyes{0}{2}]
\odot
\Gamma(\theta_{11})
\;.
\eqlabel{4a.20d}\eeq

\subsection*{4(b)When Central Matrix is a Direct Sum of D Matrices}
\mbox{}\indent
Consider first the case $\nb = 3$. Let $R(\phi) = \exp(i \sy \phi) $.
Previously we used the fact that any D matrix $D$ can be
expressed as 

\beq
D = \sum_{a, b\in Bool}
R(\phi_{ab}'') \otimes P_a \otimes P_b
\;.
\eqlabel{4b.1}\eeq
But what if $R$ were located at bit
positions 0 or 1 instead of 2? 
The next two  equations can be proven by 
expressing both sides of the equation as an $8 \times 8$ matrix.

\beq
D_0 \oplus D_1
=
\sum_{a,b\in Bool}
P_a \otimes R(\phi_{ab}') \otimes P_b
\;,
\eqlabel{4b.2}\eeq

\beq
D_{00}\oplus D_{01}\oplus D_{10}\oplus D_{11}
=
\sum_{a,b\in Bool}
P_a \otimes P_b \otimes R(\phi_{ab})
\;,
\eqlabel{4b.3}\eeq
where the $D_j$ and $D_{ij}$ are D matrices.
One can apply a string of Exchangers to move $R$ 
in Eqs.(4b.2) and (4b.3) to any bit position. Thus,

\beq
D_0 \oplus D_1
=
E(1,2) \odot
\left (
\sum_{a,b\in Bool}
 R(\phi_{ab}') \otimes P_a \otimes P_b
\right )
\;,
\eqlabel{4b.4}\eeq

\beq
D_{00}\oplus D_{01}\oplus D_{10}\oplus D_{11}
=
[E(0,1) E(1,2)]\odot
\left (
\sum_{a,b\in Bool}
R(\phi_{ab}) \otimes 
P_a \otimes  
P_b
\right )
\;.
\eqlabel{4b.5}\eeq
(Careful: $E(0, 2) \neq E(0,1) E(1,2)$. $E(0,2)$ will change
$\sum_{a,b} R(\phi_{ab})\otimes P_a \otimes P_b$
to 
$\sum_{a,b} P_b \otimes P_a \otimes R(\phi_{ab})$, which 
is not the same as the right side of Eq.(4b.3) ).

For general $\nb \geq 1$, 
if $\beta\in \bitz$ and 

\beq
E =
\left \{
\begin{array}{l}
1 \;\;\;\; {\rm if} \;\; \beta=0 \;\; {\rm or} \;\; \nb=1\\
E(\nb - \beta - 1, \nb -\beta)\cdots
E(\nb - 3, \nb - 2)
E(\nb - 2, \nb - 1)
\;\;\;\; {\rm otherwise}
\end{array}
\right .
\;,
\eqlabel{4b.6}\eeq
then 
a direct sum of $2^\beta$  D matrices
can be expressed as 

\beq
E\odot
\left (
\sum_{\va\in Bool^{\nb -1}} 
R(\phi_{\va})
\otimes P_\va
\right )
\;.
\eqlabel{4b.7}\eeq

It follows that if we want to decompose a direct sum of D matrices
into a SEO,
we can do so in 2 steps: (1) decompose 
into a SEO the D matrix that one obtains
by moving the qubit rotation to bit position $\nb -1$, (2)
Replace each bit name in the decomposition by its ``alias".
By alias we mean the new name assigned by the 
bit permutation $E$ defined by Eq.(4b.6).

\subsection*{4(c)When Central Matrix is a Diagonal Unitary Matrix}
\mbox{}\indent
Any diagonal unitary matrix $\Delta$ can be expressed as

\beq
\Delta = \sum_{\va \in Bool^{\nb}} \exp(i \phi_\va)  P_\va
\;,
\eqlabel{4c.1}\eeq
where the $\phi_\va$ are real numbers. Using the identity Eq.(2b.11) yields

\beq
\Delta =
\exp
\left (
i \sum_{ \va \in Bool^\nb }
\phi_\va
P_\va
\right )
\;.
\eqlabel{4c.2}\eeq
Now define new angles $\theta_\vb$ by

\beq
\phi_\va =
\sum_{\vb \in Bool^{\nb} }
(-1)^{\va\cdot\vb} \theta_\vb
\;.
\eqlabel{4c.3}\eeq
In terms of vectors, 

\beq
\vec{\phi} = H_{\nb} \vec{\theta}
\;,
\eqlabel{4c.4}\eeq
and

\beq
\vec{\theta} = 
\frac{1}{ 2^{\nb} }
H_{\nb} \vec{\phi}
\;.
\eqlabel{4c.5}\eeq
Let $A_\vb$ for $\vb \in Bool^\nb$ be defined by

\beq
A_\vb =
\exp
\left(
i \theta_\vb 
\sum_{\va \in Bool^{\nb}}
(-1)^{\va\cdot\vb}
P_\va
\right)
\;.
\eqlabel{4c.6}\eeq
Then $\Delta$ can be written as

\beq
\Delta =
\prod_{ \vb \in Bool^{\nb} }
A_\vb
\;,
\eqlabel{4c.7}\eeq
where the $A_\vb$ operators commute. For any $\vb \in Bool^\nb$,
we can write

\beq
\vb = \sum_{j=0}^{r-1} \vec{u}(\beta_j)
\;,
\eqlabel{4c.8}\eeq
where 

\beq
\nb-1 \geq 
\beta_{r-1} 
> \cdots
> \beta_1 > \beta_0 \geq 0
\;.
\eqlabel{4c.9}\eeq
(Careful: Compare this with Eq.(4a.11). Now
$\vb\in Bool^\nb$ instead of $Bool^{\nb -1}$ and
$\beta_{r-1}$ can be as large as $\nb-1$ instead of $\nb-2$.)
One can show using the techniques of Section 4(a) that

\beq
A_\vb
=
\left \{
\begin{array}{l}
\exp[i \theta_0] 
\;\;\;\; {\rm if} \;\;\; r=0 
\\
\exp[i \theta_\vb \sz(\beta_0)] 
\;\;\;\; {\rm if} \;\;\; r=1 
\\
\left [\cnotyes{\beta_{r-1}}{\beta_0 }
\cdots
\cnotyes{\beta_2}{\beta_0}
\cnotyes{\beta_1}{\beta_0}
\right ]\odot
\exp[ i\theta_\vb \sz(\beta_0) ]
\;\;\;\;{\rm if}\;\;\; r\geq 2
\end{array}
\right . 
\;.
\eqlabel{4c.10}\eeq
As in Section 4(a), there are other ways of decomposing $A_\vb$ into a SEO.

In conclusion, we have shown how to decompose a 
diagonal unitary matrix into a SEO. For example, suppose $\nb = 2$. Then

\beq
\Delta = diag( e^{i \phi_{00}}, e^{i \phi_{01}}, 
e^{i \phi_{10}},e^{i \phi_{11}} )
\;.
\eqlabel{4c.11}\eeq
Define $\vec{\theta}$ by 

\beq
\vec{\theta} = \frac{1}{4} H_2 \vec{\phi}
\;.
\eqlabel{4c.12}\eeq
By Eqs.(4c.7) and (4c.10),

\beq
\Delta = A_{00} A_{01} A_{10} A_{11}
\;,
\eqlabel{4c.13}\eeq
where

\beq
A_{00} = \exp( i \theta_{00} )
\;,
\eqlabel{4c.14a}\eeq

\beq
A_{01} =
I_2 \otimes
\exp( i \theta_{01} \sz )
\;,
\eqlabel{4c.14b}\eeq

\beq
A_{10} = 
\exp( i \theta_{10} \sz )
\otimes I_2 
\;,
\eqlabel{4c.14c}\eeq

\beq
A_{11} = 
\cnotyes{1}{0}
\odot
[
I_2 \otimes
\exp( i \theta_{11} \sz )
]
\;.
\eqlabel{4c.14d}\eeq

\subsection*{4(d) Comments}
\mbox{}\indent
The use of a Hadamard transform	
in Section 4(a) may seem at first somewhat mysterious
to the reader. Sorry. Here is some motivation for it.

We began Section 4(a) by noting with Eq.(4a.2) that
any D matrix $D$ can be expressed as $B = \exp[i\sy(\nb-1)]$
raised to some power which equals a linear combination
of products of $P_0(\beta) = \nbar(\beta)$ and
$P_1(\beta) = n(\beta)$, where $\beta\in Z_{0, \nb-2}$.
Using a Hadamard transform allowed us to re-express
$D$ as a product of $A_\vb$'s . According to Eq.(4a.12),
each $A_\vb$ can be expressed as $B$ raised to some power
which equals a product of $\sz(\beta)$, 
where $\beta\in Z_{0, \nb-2}$. If in Eq.(4a.2) we had
replaced each $n(\beta)$ and $\nbar(\beta)$
by $\frac{1}{2}[1-\sz(\beta)]$ and 
$\frac{1}{2}[1+\sz(\beta)]$, respectively,
then we would have found the same
thing that we found via a Hadamard transform---i.e., 
that $D$ is expressible as a product of $A_\vb$'s, each of
which is expressible as $B$ raised to some
product of $\sz(\beta)$'s.

For any $2 \times 2$ complex matrix $B$ such that $\det(B)\neq 0$,
we define a {\it controlled gate} $g$ to
be an operator of the form

\beq
g = B(\alpha)^{P_{b_0}(\beta_0)P_{b_1}(\beta_1)\ldots P_{b_{r-1}}(\beta_{r-1})}
\;,
\eqlabel{4d.1}\eeq
where $b_o, b_1, \ldots, b_{r-1}\in Bool$, and 
$\beta_0, \beta_1, \ldots, \beta_{r-1}, \alpha$ are distinct elements of $\bitz$.
We call $\beta_0, \beta_1, \ldots, \beta_{r-1}$
the {\it control bits} and $\alpha$ the {\it flipper bit}.
 An example of Eq.(4d.1) is when $B = e^{i\phi}$
for some real $\phi$, in which case we call $g$ a {\it controlled phase factor}.
We already encountered controlled phase factors, with 2 controls,
in the FFT algorithm. Two other examples of Eq.(4d.1) are when $B = \sx(\alpha)$ 
and $B = \exp[i  \vec{\theta}\cdot\vec{\sigma}(\alpha)]$ 
for some real 3-dimensional vector
$\vec{\theta}$, in which cases we get a {\it c-not} (with possibly more than
one control) and a {\it controlled qubit rotation}.

In general, whenever one has a controlled gate $g$
with more than one control bit,
one can simplify $g$
by expressing its exponents 
$P_0(\beta) = \nbar(\beta)$ and
$P_1(\beta) = n(\beta)$ in terms of 
$I_2(\beta)$ and $\sz(\beta)$.
Appendices C and D say more about this transformation.
Appendix D discusses the transformation from
the point of view of Linear Algebra.
Appendix C uses the transformation to decompose
a controlled gate with 2 controls into a SEO.
Decompositions like the one in Appendix C 
were first discovered by the authors of Ref.\enote{CastOfThousands}.
However, their method of deriving such decompositions
is very different from ours.

\section*{5. SEO-Length Optimizations}
\mbox{}\indent	
In previous sections, we have given an algorithm that 
can decompose an arbitrary unitary matrix into a SEO.
But the SEO's generated by this algorithm may not be
the shortest possible. In this chapter we will 
indicate several ``SEO-length optimizations"---small
adjustments to the algorithm which are either
guaranteed or, at least, likely to produce
shorter SEO's.

The group $U(\ns)$ of $\ns \times \ns$ unitary
matrices  has $\ns^2$ free (real) parameters. (It has
$\ns^2$ complex entries for a total of $2 \ns^2$
real parameters, but those parameters must satisfy 
$\ns^2$ orthonormality constraints.) In a 
CSD tree, level $r$ has $2^{r-1}$ nodes so the number of 
nodes in the tree is 
$1 + 2 + 2^2 +\ldots + 2^\nb = 2^{\nb + 1} -1 \approx 2 \ns$.
(Level $\nb +1$ alone has $\ns$ of the nodes, about half of them!)
Each node yields Order($\ns$) elementary operations.
Therefore, in general, a CSD tree yields a SEO 
whose length is Order($\ns^2$). In essence, each elementary
operation of the SEO (except permutation operations like c-nots) 
carries (approximately) one of the $\ns^2$ free parameters.

But we know that certain families of  matrices contained in $U(\ns)$
have internal symmetries which allow us to 
parametrize them with substantially fewer than $\ns^2$ parameters.
And we know how to decompose some of these families
(for example, the DFT matrices) into a SEO whose length is a polynomial
in $\nb$. Can such SEOs be obtained with our CSD tree algorithm?
As we shall see, the answer is yes, at least in some important cases
like the DFT matrices. The trick is to introduce optimizations into 
our algorithm which make the CSD tree degenerate into a 
simple string (or nearly one) of nodes. The other nodes 
never ``grow".  For any node $\caln$, if its 
right side matrices(i.e., $R(\vb, \caln)$ in Fig.1,
where $\vb\in Bool^\lambda$ and $\lambda$ is the level of $\caln$) are all equal
to the identity matrix, then 
$\caln$ bears no children on its right side. Ditto for 
its left side matrices. Note that in Fig.2,
the first (ditto, second) quantum FFT expansion looks like what one 
would expect if a CSD tree were to degenerate into the string 
consisting of the rightmost (ditto, leftmost) node of each level of the tree. 
As we shall see, this is precisely what happens if we add
the following optimizations to our
CSD tree algorithm.

\subsection*{5(a)Lazy Ordering of Factors}
\mbox{}\indent

Consider Eq.(4a.19). The operators $A_\vb$ commute so
this equation is valid regardless of the order in 
which the $A_\vb$  are multiplied.
Suppose that we multiply the $A_\vb$ so that 
their subscripts $\vb$ are in a lazy ordering 
that starts with 00 on the right and ends with
01 on the left:

\beq
D = A_{01} A_{11} A_{10} A_{00}
\;.
\eqlabel{5a.1}\eeq
After inserting Eqs.(4a.20) for the $A_\vb$, this yields

\beq
D = 
\cnotyes{0}{2}\Gamma(\theta_{01})
\cnotyes{1}{2}\Gamma(\theta_{11})
\cnotyes{0}{2}\Gamma(\theta_{10})
\cnotyes{1}{2}\Gamma(\theta_{00})
\;.
\eqlabel{5a.2}\eeq
We see that by ordering the $A_\vb$ in this way,
several c-nots cancel out. Only one c-not remains
between adjacent $\Gamma$'s. There is no c-not
to the right of the rightmost $\Gamma$ in Eq.(5a.2),
because we started the lazy ordering with $\vb =00$.
There is only one c-not to the left of the leftmost
$\Gamma$ in Eq.(5a.2), because we ended the lazy ordering 
with $\vb = 01$.

The above example assumes $\nb = 3$ and that we are
decomposing a central matrix of type 1 (i.e., a central 
matrix which is a single D matrix). However,
this method of shortening a SEO can also be used 
for the other two types
of central matrices and for other values of $\nb$. Indeed, the other two types
of central matrices are also decomposed by the algorithm
into a product of $A_\vb$ type matrices. Furthermore,
it is possible to find for arbitrary $\nb$
a lazy ordering of $Bool^\nb$ which starts with 
the zero vector $\vb = 0$ and ends with a $\vb$
which has only one non-zero component.

\subsection*{5(b)Lightening Right Side Matrices}
\mbox{}\indent
Suppose $D$ is a $4\times 4$ D matrix such that its two
angles are equal. Hence,

\beq
D = 
\left[
\begin{array}{cc}
c I_2 & s I_2 \\
-s I_2 & c I_2 
\end{array}
\right]
\;,
\eqlabel{5b.1}\eeq
where $c = \cos \theta, s = \sin \theta$ for some angle $\theta$.
Then for any $2\times 2$ unitary matrix $G_0$,

\beq
(G_0 \oplus G_0)
D 
(G_0^{\dagger} \oplus G_0^{\dagger})
= D
\;.
\eqlabel{5b.2}\eeq

More generally, consider an $N \times N$ D matrix $D$. 
Let $\theta_a$ for $a\in Z_{0, \frac{N}{2} -1 }$
be its angles, arranged in non-decreasing order (recall Appendix A). 
Suppose the first $d_0$
angles are equal to each other, the next $d_1$ angles are
equal to each other but larger than the previous $d_1$ angles,
and so on. Hence,
$d_0 + d_1 +\ldots +d_{M -1} = \frac{N}{2}$.
We will call $d_0, d_1, \ldots, d_{M -1}$ the {\it degeneracies 
of the D matrix angles}. For $a\in Z_{0, M -1 }$, let $G_a$ be
any $d_a \times d_a$ unitary matrix, and define $G$ by

\beq
G = G_0 \oplus G_1 \oplus \ldots \oplus G_{M-1}
\;.
\eqlabel{5b.3}\eeq
Then

\beq
(G \oplus G)
D 
(G^{\dagger} \oplus G^{\dagger})
= D
\;.
\eqlabel{5b.4}\eeq

Suppose that 

\beq
U = (L_0 \oplus L_1) D (R_0 \oplus R_1) 
\;
\eqlabel{5b.5}\eeq
is a CSD of a unitary matrix $U$.
If $G$ is a unitary matrix that satisfies Eq.(5b.4), 
define $L'_j$ and $R'_j$ for $j\in Bool$ by

\beq
L'_j = L_j G
\;,
\eqlabel{5b.6}\eeq

\beq
R'_j = G^{\dagger} R_j
\;.
\eqlabel{5b.7}\eeq
Then 

\beq
U = (L'_0 \oplus L'_1) D (R'_0 \oplus R'_1) 
\;
\eqlabel{5b.8}\eeq
is also a CSD of $U$.
Suppose $D$ has $M$ distinct angles with degeneracies 
$d_0, d_1, \ldots d_{M-1}$. Define $\Sigma_{-1} = 0$ and 
$\Sigma_a = d_0 + d_1 + \cdots + d_a$ for all $a\in Z_{0, M-1}$.
  For $a\in Z_{0, M-1}$ and 
$j\in Bool$, suppose $R'^{(a)}_j$ (ditto, $R^{(a)}_j$)
are the $d_a$ rows with row indices from $\Sigma_{a-1}$ to $\Sigma_a -1$ 
of $R'_j$ (ditto, $R_j$). Then
Eq.(5b.7) is equivalent to the equations

\beq
R^{(a)}_j = G_a R'^{(a)}_j
\;,
\eqlabel{5b.9}\eeq
for $a\in Z_{0, M-1}$ and 
$j\in Bool$.
For each $a\in Z_{0, M-1}$, we will choose the 
two matrices $G_a$ and  $R'^{(a)}_0$
on the right side of the last equation
so that they are the $Q$ and $R$ matrices, respectively,
in a QR decomposition of $R^{(a)}_0$. In a QR
decomposition\enote{Noble}, the Q matrix
is unitary so constructing $G_a$ this way
will make it unitary, as required by our original definition of 
it.  For any rectangular matrix M, its 
principal diagonal are those entries of M for which the
row and column indices are equal.
In a QR decomposition, the R matrix 
has zero entries below the principal diagonal so 
$R'^{(a)}_0$ will have this property.
In a QR decomposition, it is
possible to multiply each row of the R matrix by a phase factor and
the corresponding column of Q by the conjugate phase factor.
Thus it is possible
for each $a\in Z_{0, M-1}$ to do
 a QR decomposition 
in which all the terms along the principal diagonal of 
$R'^{(a)}_0$ are non-negative. (Alternatively, one can make non-negative
all the terms along the principal diagonal of $R'_0$.)
 
 By  {\it lightening the right side matrices}, we will mean 
 replacing the CSD Eq.(5b.5) by 
 the CSD Eq.(5b.8), where $G$ is chosen so that
 for all $a\in Z_{0, M-1}$,
 the $R'^{(a)}_0$ have zero entries below the principal diagonal
 and they (or else $R'_0$) all have non-negative diagonal entries.
 The strategy of lightening the right side matrices
 is not guaranteed to make a CSD tree have fewer nodes.
 However, whenever it does achieve this goal,
 the reason why it does so is clear. The strategy
 tries to make zero as many entries of $R'_0$ as possible.
 This causes the CSD tree to grow mostly to the left. Its 
 right side growth is stunted.
 
 \subsection*{5(c) Extracting Phases From Complex D Matrices}
\mbox{}\indent
For this optimization, it is convenient to generalize
the definitions of a D matrix and of a central matrix.
Henceforth, let a {\it D matrix} be any unitary matrix $D$
which can be written in the form

\beq
D =
\left[
\begin{array}{cc}
D_{00} & D_{01}\\
D_{10} & D_{11}
\end{array}
\right]
\;,
\eqlabel{5c.1a}\eeq
where the $D_{i,j}$ for $i, j \in Bool$ are same-sized
diagonal matrices. An equivalent definition: a D matrix
is a matrix of the form

\beq
D = \sum_{\va \in Bool^{\nb-1}} U^{(\va)}\otimes P_{\va}
\;,
\eqlabel{5c.1b}\eeq
where the $U^{(\va)}$ are unitary $2\times 2$ matrices.
What we previously called a D matrix 
will now be called a {\it real D matrix}. If we need to emphasize
that a D matrix is of the new kind, 
we will describe it as a {\it complex D matrix}.
Henceforth, let a {\it central matrix} be defined 
as before, except that the term ``D matrix" in the 
previous definition of a central matrix should now be interpreted
to mean a complex D matrix.
What we previously called a central matrix 
will now be called a {\it bare central matrix}(we can't call it 
a real central matrix because it could be a 
complex unitary diagonal matrix).
If we need to emphasize
that a central  matrix is of the new kind,
we will describe it as a {\it dressed central matrix}.

In previous sections, we showed that: (1) Given
an arbitrary unitary matrix $U$, one can construct
a CSD tree in which each node stores a bare central 
matrix such that $U$ equals the product (in a particular order)
of these central matrices. (2) Any bare central
matrix can be decomposed into a SEO.
In this section, we will show (1) and (2),
but with the words ``bare central matrix" replaced 
by ``dressed central matrix".

To construct a CSD tree of dressed central matrices,
one proceeds as follows. Suppose node $\caln$
at level $\lambda \in Z_{1, \nb-1}$
has children nodes $\caln_L$ and $\caln_R$
to its left and right, respectively.
Node $\caln$ will store a dressed central matrix $Q(\caln)$,
a collection of left side matrices $L(\vb, \caln)$
for $ \vb \in Bool^\lambda$, and 
a collection of right side matrices $R(\vb, \caln)$
for $ \vb \in Bool^\lambda$.
One performs a normal CSD only on  those side matrices $R(\vb, \caln)$
and $L(\vb, \caln)$ which are not complex D matrices.
If a side matrix is a complex D matrix, one performs an aborted 
CSD. Given a subroutine $f(U, L_0, L_1, D, R_0, R_1)$ 
for doing CSD, with input $U$ and outputs  
$L_0, L_1, D, R_0, R_1$, an {\it aborted CSD} 
is defined as a pass through the subroutine that 
returns the identity matrix for the side matrices 
$L_0, L_1, R_0, R_1$ and it returns $U$ for $D$.

So far we have shown (1)---
how  to construct a CSD tree of dressed 
central matrices. Next we will show (2)---
how to decompose a dressed central matrix into a SEO.
We will show how any dressed central matrix $Q$
can be factored into 

\beq
Q = \Delta_L Q' \Delta_R
\;,
\eqlabel{5c.2}\eeq
where $Q'$ is a bare central matrix with real entries
and where $\Delta_L$ and $\Delta_R$
are diagonal unitary matrices (so they are also 
bare central matrices, though not necessarily real).
In previous sections, we showed how to decompose
any bare central matrix into a SEO. So we 
will be done proving (2) if we can show Eq.(5c.2)

Consider a $2\times 2$ complex D matrix $D$.
One can always express $D$ as 

\beq
D = 
\left[
\begin{array}{cc}
c e^{i\Omega} & s e^{i(\Omega + \omega_R)} \\
-s e^{i(\Omega + \omega_L)} & c e^{i(\Omega + \omega_L + \omega_R)}
\end{array}
\right]
\;,
\eqlabel{5c.3}\eeq
where $\Omega, \omega_L, \omega_R, c, s$
are real numbers, and where 
$c^2 + s^2 = 1$.
Eq.(5c.3) implies

\beq
D = 
diag(1, e^{i \omega_L})
\left[
\begin{array}{cc}
c & s \\
-s & c
\end{array}
\right]
e^{i \Omega}
diag(1, e^{i \omega_R})
\;.
\eqlabel{5c.4}\eeq
Note that we can always choose
$s\geq 0$. For if 
$s<0$, then one can replace 
$s\rarrow |s|$, 
$\omega_L \rarrow \omega_L + \pi$ and 
$\omega_R \rarrow \omega_R + \pi$.
Likewise, we can always choose $c \geq 0$.
For if 
$c<0$, then one can replace 
$c\rarrow |c|$, 
$\omega_L \rarrow \omega_L + \pi$,
$\omega_R \rarrow \omega_R + \pi$ and
$\Omega \rarrow \Omega + \pi$.
With $s\geq 0$ and $c\geq 0$, we can 
write $c = \cos \theta$ and $s = \sin \theta$
with $\theta \in [0, 90^0]$. It is a trivial
exercise to solve for the parameters 
$\Omega, \omega_L, \omega_R$ and $\theta$
in terms of the entries of $D$.

Now consider an $N\times N$ complex D matrix $D$.
Let

\beq
\Gamma_\sigma = diag( 
e^{i \omega_\sigma^{(0)}}
e^{i \omega_\sigma^{(1)}},
\ldots,
e^{i \omega_\sigma^{(\frac{N}{2}-1)}}
)
\;\;\;\;{\rm for}\;\; \sigma \in \{L, R\}
\;,
\eqlabel{5c.5}\eeq

\beq
\Gamma = diag( 
e^{i \Omega^{(0)}}
e^{i \Omega^{(1)}},
\ldots,
e^{i \Omega^{(\frac{N}{2}-1)}}
)
\;,
\eqlabel{5c.6}\eeq

\beq
C = diag(C_0, C_1, \ldots, C_{\frac{N}{2}-1})
\;,
\eqlabel{5c.7}\eeq

\beq
S = diag(S_0, S_1, \ldots, S_{\frac{N}{2}-1})
\;,
\eqlabel{5c.8}\eeq
where for any $j \in Z_{0,\frac{N}{2}-1}$
and any $\sigma \in \{L, R\}$, 
$\Omega^{(j)}$ and 
$\omega_\sigma^{(j)}$ are real, and
$C_j = \cos\theta_j$, 
$S_j = \sin\theta_j$ with 
$\theta_j \in [0, 90^0]$.
One can always express $D$ as

\beq
D = [I_{\frac{N}{2}} \oplus \Gamma_L]
\left[
\begin{array}{cc}
C & S \\
-S & C
\end{array}
\right]
[\Gamma \oplus (\Gamma \Gamma_R)]
\;.
\eqlabel{5c.9}\eeq
Note that this is really a special CSD.
To harmonize with 
our similar CSD angle convention (see Appendix A),
it is important that we choose the angles 
$\theta_j$ to be in the interval $[0, 90^0]$.
As in the $N=2$ case, it is trivial to solve
for the parameters 
$\Omega^{(j)}$,
$\omega_\sigma^{(j)}$ and 
$\theta_j$ in terms of the entries of $D$.

Now consider any dressed central matrix $Q$. We want to show that
$Q = \Delta_L Q' \Delta_R$. This equation is trivially satisfied if $Q$ is
a diagonal unitary matrix so assume instead that  $Q$  is 
a direct sum of one or more complex D matrices.
For example, suppose that

\beq
Q = D(00) \oplus D(01) \oplus D(10) \oplus D(11)
\;,
\eqlabel{5c.10}\eeq
where the matrices $D(\vb)$ for $\vb \in Bool^2$ are
complex D matrices. By Eq.(5c.9), for each $\vb \in Bool^2$,
one can write 

\beq
D(\vb) = \Delta_L(\vb) D'(\vb) \Delta_R(\vb)
\;,
\eqlabel{5c.11}\eeq
where the matrices 
$\Delta_L(\vb)$ and $\Delta_R(\vb)$
are diagonal unitary matrices and 
$ D'(\vb)$ is a real D matrix.
$Q = \Delta_L Q' \Delta_R$ now follows if we set

\beq
\Delta_\sigma = 
\Delta_\sigma(00) \oplus
\Delta_\sigma(01) \oplus
\Delta_\sigma(10) \oplus
\Delta_\sigma(11)
\;,
\eqlabel{5c.12}\eeq
for $\sigma \in \{L, R\}$ and

\beq
Q' = 
D'(00) \oplus
D'(01) \oplus
D'(10) \oplus
D'(11)
\;.
\eqlabel{5c.13}\eeq

One can expand $\Delta_L$, $Q'$ and $\Delta_R$
into a SEO using the techniques of Section 4. 
Alternatively, one might do this only
for $Q'$, and express $\Delta_L$ and $\Delta_R$
as a product of controlled phase factors.
We end this section with a discussion of
controlled phase factors and of their products.

Any $\ns \times \ns$ 
diagonal unitary matrix $\Delta$ can 
be expressed as a product of controlled phase factors. For 
example, suppose 

\beq
\Delta = diag(
e^{i\phi_{00}},
e^{i\phi_{01}},
e^{i\phi_{10}},
e^{i\phi_{11}})
\;,
\eqlabel{5c.14}\eeq
where  $\phi_\vb$  is a
real number for all $\vb\in Bool^2$.  Eq.(5c.14) can be rewritten as

\beq
\Delta = 
e^{i\phi_{00}\nbar(1)\nbar(0)}
e^{i\phi_{01}\nbar(1)n(0)}
e^{i\phi_{10}n(1)\nbar(0)}
e^{i\phi_{11}n(1)n(0)}
\;.
\eqlabel{5c.15}\eeq
By replacing
all  exponents $\nbar(\beta)$
with 
$ 1 - n(\beta)$, the last equation
can be expressed in the form:

\beq
\Delta =
e^{i\theta_{00}}
e^{i\theta_{01}n(0)}
e^{i\theta_{10}n(1)}
e^{i\theta_{11}n(1)n(0)}
\;,
\eqlabel{5c.16}\eeq
where the $\theta_\vb$ are real numbers,
linear combinations of the $\phi_\vb$.
The SEO of Eq.(5c.16) is preferable to
that of Eq.(5c.15) because Eq.(5c.15)
contains four 2-bit operations whereas 
Eq.(5c.16) contains only one. 
From the point of view of Linear Algebra,
going from Eq.(5c.15) to Eq.(5c.16)
is an example of
changing from the ${\cal B}(n, \nbar)$
to the ${\cal B}(n, 1)$ basis.
See Appendix E
for more information about this transformation.

In this paper, we use bare central matrices whose D matrices have angles
contained in the interval $[0, 90^o]$.
An interesting special case is when these angles are all
either 0 or $90^o$ exclusively. 
The entries of such central matrices are elements
of $\{-1, 0, 1\}$. When the optimization being
discussed in this section is turned ON, 
it is sometimes convenient to decompose such central matrices
into a SEO by a method different
from the one discussed in Section 4. See Appendix F for more 
information about this.

\subsection*{5(d)Permuting Bits before Each CSD}
\mbox{}\indent
This optimization is only partially implemented in 
the current version of Qubiter, and it was not used
to get any of the results of Section 6.

		\begin{center}
			\epsfig{file=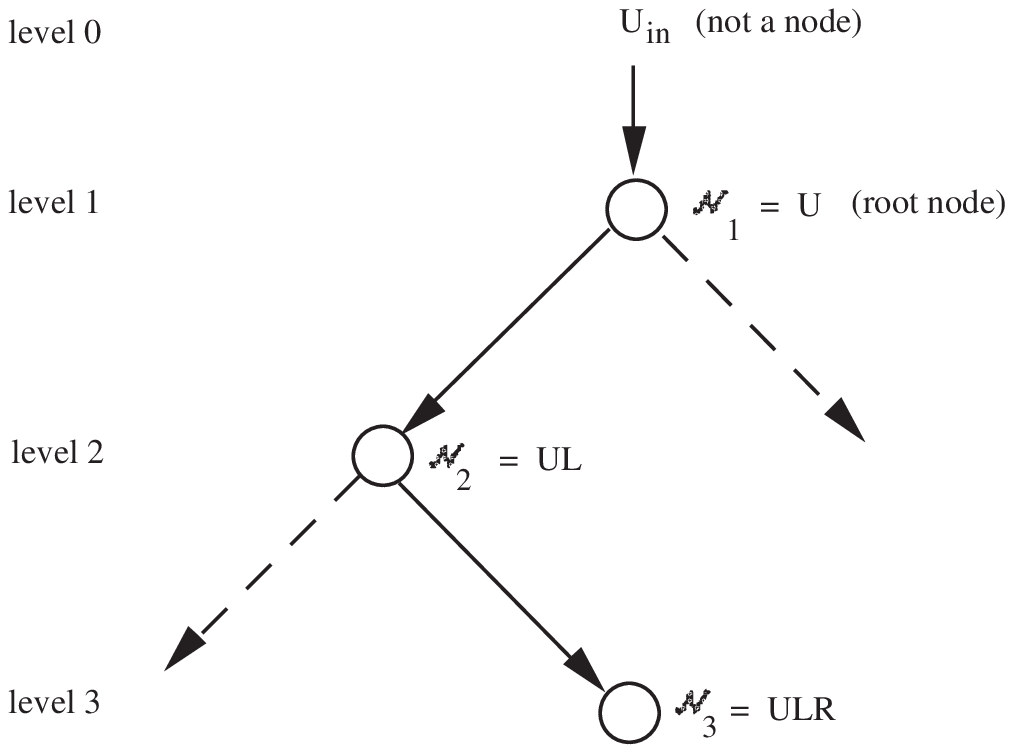}
			
			{Fig.8 Subset of CSD tree.}
		\end{center}

We will start with an example and then generalize it. Suppose
that $\nb\geq 3$ and that we are interested in finding
the central matrix belonging to node $\caln_3 = ULR$
of Fig.8. This being a binary tree, $\caln_3$
has only one parent, call it $\caln_2 = UL$. 
$\caln_2$ in turn has just one parent $\caln_1 = U$, which is 
the level 1 node (root node) of Fig.8.
If $U_{in}$ is the initial unitary matrix and $P(0, \caln_1)$
is some permutation matrix that 
permutes bits (hence, $P(0, \caln_1)$ is
always expressible as a product of Exchangers),
we do a CSD of $P(0, \caln_1)\odot U_{in}$
instead of $U_{in}$:

\beq
P(0, \caln_1)\odot U_{in} =
[L(0, \caln_1) \oplus L(1, \caln_1)]
D(0, \caln_1)
[R(0, \caln_1) \oplus R(1, \caln_1)]
\;.
\eqlabel{5d.1a}\eeq
To get the central matrix belonging to $\caln_2$, we 
permute the bits of $L(0, \caln_1)$ and
$L(1, \caln_1)$, and then do a CSD of the outcome:

\beq
P(a, \caln_2) \odot L(a, \caln_1) =
[L(a0, \caln_2) \oplus L(a1, \caln_2)]
D(a, \caln_2)
[R(a0, \caln_2) \oplus R(a1, \caln_2)]
\;,
\eqlabel{5d.1b}\eeq
for all $a\in Bool$. Finally, to get the central matrix
belonging to $\caln_3$, we permute the bits of 
$R(ab, \caln_2)$ for all $a,b\in Bool$, and then
do a CSD of the outcome:

\beq
P(ab, \caln_3) \odot R(ab, \caln_2) =
[L(ab0, \caln_3) \oplus L(ab1, \caln_3)]
D(ab, \caln_3)
[R(ab0, \caln_3) \oplus R(ab1, \caln_3)]
\;,
\eqlabel{5d.1c}\eeq
for all $a,b \in Bool$. The central matrix $Q(\caln_j)$
for $j\in Z_{1,3}$ is given by

\beq
Q(\caln_1) = D(0, \caln_1)
\;,
\eqlabel{5d.2a}\eeq

\beq
Q(\caln_2) = D(0, \caln_2) \oplus D(1, \caln_2)
\;,
\eqlabel{5d.2b}\eeq

\beq
Q(\caln_3) = 
D(00, \caln_3) \oplus
D(01, \caln_3) \oplus
D(10, \caln_3) \oplus
D(11, \caln_3)
\;.
\eqlabel{5d.2c}\eeq
One can define a permutation matrix $P(\caln_j)$
for $j\in Z_{1,3}$ by

\beq
P(\caln_1) = P(0, \caln_1)
\;,
\eqlabel{5d.3a}\eeq

\beq
P(\caln_2) = P(0, \caln_2) \oplus P(1, \caln_2)
\;,
\eqlabel{5d.3b}\eeq

\beq
P(\caln_3) = 
P(00, \caln_3) \oplus
P(01, \caln_3) \oplus
P(10, \caln_3) \oplus
P(11, \caln_3)
\;.
\eqlabel{5d.3c}\eeq
It is useful to define matrices $Q'(\caln_j)$
for $j\in Z_{1,3}$
by undoing the permutations that were performed 
to calculate $Q'(\caln_j)$:

\beq
Q'(\caln_1) = P^{-1}(\caln_1) \odot Q(\caln_1)
\;,
\eqlabel{5d.4a}\eeq

\beq
Q'(\caln_2) = [P(\caln_2)P(\caln_1)]^{-1} \odot Q(\caln_2)
\;,
\eqlabel{5d.4b}\eeq

\beq
Q'(\caln_3) = [P(\caln_3)P(\caln_2)P(\caln_1)]^{-1} \odot Q(\caln_3)
\;.
\eqlabel{5d.4c}\eeq

When the optimization presently being discussed is turned OFF,
we associate with each node $\caln$
a central matrix $Q(\caln)$, and the product of these $Q(\caln)$ 
in some special order equals $U_{in}$. With this
optimization turned ON, we associate with each $\caln$
a pair $(Q(\caln), P(\caln))$ consisting of
a central matrix $Q(\caln)$ and a permutation matrix
$P(\caln)$, and  
the product of the $Q'(\caln)$ 
equals $U_{in}$.
Note that the matrices $Q'(\caln)$
are not necessarily central matrices.

The above discussion can be generalized to arbitrary $\nb$ as follows.
Assume $\caln_\lambda$ is a node at level $\lambda \in Z_{1, \nb}$.
$\caln_\lambda$ has a string of predecessors 
$\caln_{\lambda -1}, \caln_{\lambda - 2}, \ldots \caln_1$,
where $\caln_{j-1}$ is the single parent of $\caln_j$ and 
$\caln_1$ is the tree's root node. Define $Bool^0 =\{0\}$.
If $\lambda = 1$, let $U_{\lambda-1} = U_{in}$.
If $\lambda > 1$, let $U_{\lambda-1}$
equal either 
$R(\vb , \caln_{\lambda-1})$
or
$L(\vb , \caln_{\lambda-1})$
for some
$\vb \in Bool^{\lambda -1}$.
For any $\vb\in Bool^{\lambda-1}$,
the $D, R, L$ matrices of node $\caln_\lambda$
are obtained by applying to $U_{\lambda-1}$
a permutation matrix $P(\vb, \caln_\lambda)$
that permutes its bits
and then doing a CSD on the outcome:
 
\beq
P(\vb, \caln_\lambda)\odot U_{\lambda-1}=
[L(\vb 0, \caln_\lambda) \oplus L(\vb 1, \caln_\lambda)]
D(\vb, \caln_\lambda)
[R(\vb 0 , \caln_\lambda) \oplus R(\vb 1, \caln_\lambda)]
\;.
\eqlabel{5d.5}\eeq
Define the central matrix $Q(\caln_\lambda)$ of node 
$\caln_\lambda$ by

\beq
Q(\caln_\lambda) = \bigoplus_{\vb \in Bool^{\lambda -1}} D(\vb, \caln_\lambda)
\;.
\eqlabel{5d.6}\eeq
Define the permutation matrix $P(\caln_\lambda)$ of node 
$\caln_\lambda$ by

\beq
P(\caln_\lambda) = \bigoplus_{\vb \in Bool^{\lambda-1}} P(\vb, \caln_\lambda)
\;.
\eqlabel{5d.7}\eeq
Finally, define the matrix $Q'(\caln_\lambda)$ of 
node $\caln_\lambda$ by

\beq
Q'(\caln_\lambda) =
[P(\caln_\lambda)\ldots P(\caln_2) P(\caln_1)]^{-1}\odot Q(\caln_\lambda)
\;.
\eqlabel{5d.8}\eeq

With this optimization ON, we associate a pair
$(Q(\caln), P(\caln))$ with each node $\caln$
of the CSD tree, and the product of 
the (not necessarily central) matrices $Q'(\caln)$
in some order equals $U_{in}$.

How to best choose the permutation matrices $P(\caln)$
for this optimization remains an open question.
One can carry out two main types of permutation searches: 
exhaustive or heuristic. An example of
a partially exhaustive permutation search (and the only 
permutation search implemented in the current 
version of Qubiter) is as follows.
Try all possible bit permutations $P(\caln_1)$
for the root node. Set $P(\caln)$ equal 
to the identity for all other nodes. 
Of all possibilities tried, find that which
yields the shortest SEO. Obviously,
one can try more complicated kinds of 
exhaustive searches.
As for heuristic searches,
for 
each node $\caln$ at level $\lambda$,
one could try to find a permutation $P(\caln)$ 
which produces the smallest
possible number of distinct CSD angles in the matrices $D(\vb, \caln)$
for all $\vb \in Bool ^{\lambda-1}$.
One expects that increasing the degeneracy in these CSD angles
will make 
the ``lightening right-side matrices" optimization
more effective.

\section*{6. Qubiter}
\mbox{}\indent	
At present, Qubiter is a very 
rudimentary program. The current version--- 
Qubiter1.1---is  written in pure C++, and has no graphical user interface.
All optimizations discussed in Section 5 except
5(d) have been fully implemented in Qubiter1.1.
In its ``compiling" mode, Qubiter takes as input
a file with the entries of a unitary matrix and returns as output
a file with a SEO. In its ``decompiling"
mode, it does the opposite: it takes a SEO file and 
returns the entries of a matrix.
The lines in a SEO file are of 
6 types.
\begin{description}

\item{(a) ROTY \qquad $\alpha$ \qquad $ang$}\newline 
where $\alpha \in \bitz$ and $ang$ is a real number. This
signifies the rotation of qubit $\alpha$ about
 the Y axis by an angle of $ang$ in degrees.
In other words, $\exp(i\sy(\alpha)ang\frac{\pi}{180})$.
(Some people would call this a rotation by $2\;\;ang$ instead of $ang$).

\item{(b) ROTZ \qquad $\alpha$ \qquad $ang$}\newline 
This is the same as (a) except that the rotation is about
the Z axis instead of the Y one.

\item{(c) SIGX \qquad $\alpha$}\newline 
where $\alpha \in \bitz$. This
signifies unconditional flipping (NOT)  of qubit $\alpha$.
In other words, $\sigma_x(\alpha)$. 

 \item{(d) CNOT 
\quad $\alpha_1$ \quad $char_1$
\quad $\alpha_2$ \quad $char_2$
\quad $\ldots$
\quad $\alpha_{r}$ \quad $char_{r}$
\quad $\beta$}\newline 
where $r\geq 1$, $\alpha_1, \alpha_2, \dots \alpha_{r},\beta$ are distinct
elements of $\bitz$ and
$char_1, char_2, \dots char_{r}$ are elements of $\{T, F\}$.
This signifies a controlled-not with $r$ controls.
First suppose $r=1$. 
If $char_1$ is the character $T$, this signifies $\cnotyes{\alpha_1}{\beta}$.
Read it as: ``c-not: if $\alpha_1$ is true, then flip $\beta$."
If $char_1$ is the character $F$, this signifies $\cnotno{\alpha_1}{\beta}$.
Read it as: ``c-not: if $\alpha_1$ is false, then flip $\beta$."
Cases with $r>1$ are defined analogously. 
For example, CNOT 0 T 1 F 2 signifies 
$\sx(2)^{n(0)\nbar(1)}$.
Read it as: ``c-not: if bit 0 is true
and bit 1 is false, then flip bit 2."

\item{(e) PHAS \qquad $ang$}\newline 
where $ang$ is a real number. This signifies 
a phase factor $\exp(i(ang)\frac{\pi}{180})$.
Thus, $ang$ is an angle expressed in degrees.

\item{(f) CPHA 
\quad $\alpha_1$ \quad $char_1$
\quad $\alpha_2$ \quad $char_2$
\quad $\ldots$
\quad $\alpha_{r}$ \quad $char_{r}$
\quad $ang$}\newline 
where $r\geq 1$, $\alpha_1, \alpha_2, \dots \alpha_{r}$ are distinct
elements of $\bitz$,
$char_1, char_2, \dots char_{r}$ are elements of $\{T, F\}$,
and $ang$ is a real number. This signifies a controlled phase factor
with $r$ controls.
For example, PHAS 0 T 1 F 90 would signify 
$[\exp(i 90 \frac{\pi}{180})]^{n(0)\nbar(1)}$.
Read it as: ``phase shift: if bit 0 is true
and bit 1 is false, then shift phase by 90 degrees."
Thus, $ang$ is an angle expressed in degrees.
\end{description} 
CNOT with more than one control and CPHA with more than two
are not elementary; i.e., they act on more than 2 bits.
Strictly speaking, non-elementary operations should not be allowed in a SEO.
However, Qubiter allows these 2 types because they are popular in the literature.
We have indicated in previous sections
how to decompose such controlled gates into SEOs. 

Fig.9 shows the output of Qubiter for the 2, 3 and 4 
bit Hadamard matrices, with all optimizations mention in Section
5 except 5(d) turned ON. Note that Qubiter is smart enough to realize
that it is dealing here with a tensor product of 1 bit matrices.
Fig.10 shows the output of Qubiter for the 2, 3 and 4 
bit DFT matrices, with all optimizations mentioned in Section
5 except 5(d) turned ON. Note that the quantum FFT algorithm
 of Ref.\enote{Copper} is 
exactly reproduced. Thus, the quantum FFT algorithm is a special case of 
our CSD tree algorithm.
In both the Hadamard and DFT cases, the CSD tree 
degenerates into the string  consisting
of the leftmost node of each level of the tree.

In the future, we plan to introduce into Qubiter more 
optimizations and some quantum error correction. 
Much work remains to be done.

		\begin{center}
			\epsfig{file=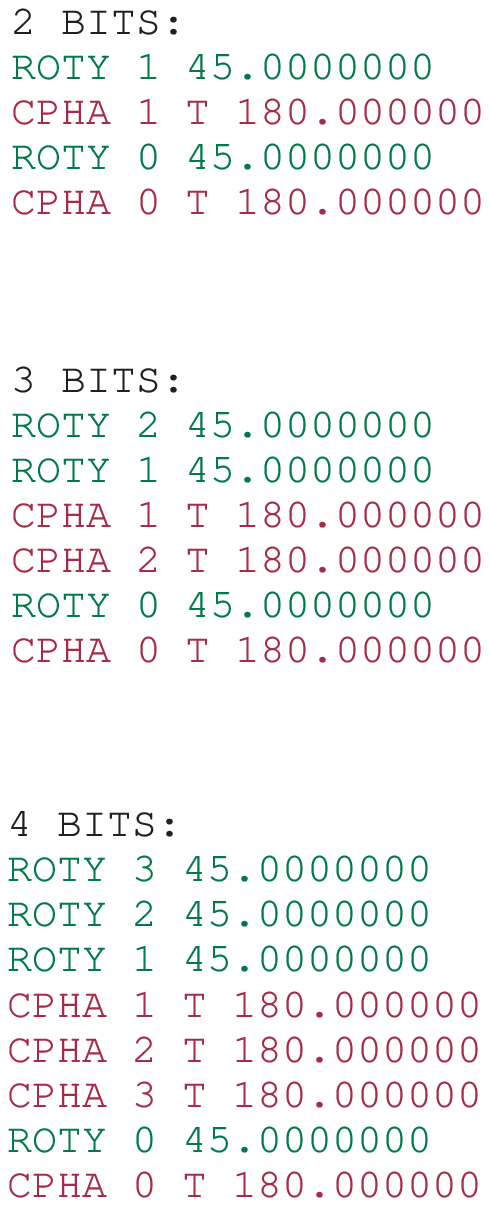}
			
			{Fig.9 Output of Qubiter with input matrix equal to 
			$\frac{1}{\sqrt{\ns}} H_{\nb}$ for $\nb = 2, 3, 4$.
			$H_{\nb}$ is the $\nb$ bit Hadamard matrix defined in Section 2(c).}
		\end{center}

		\begin{center}
			\epsfig{file=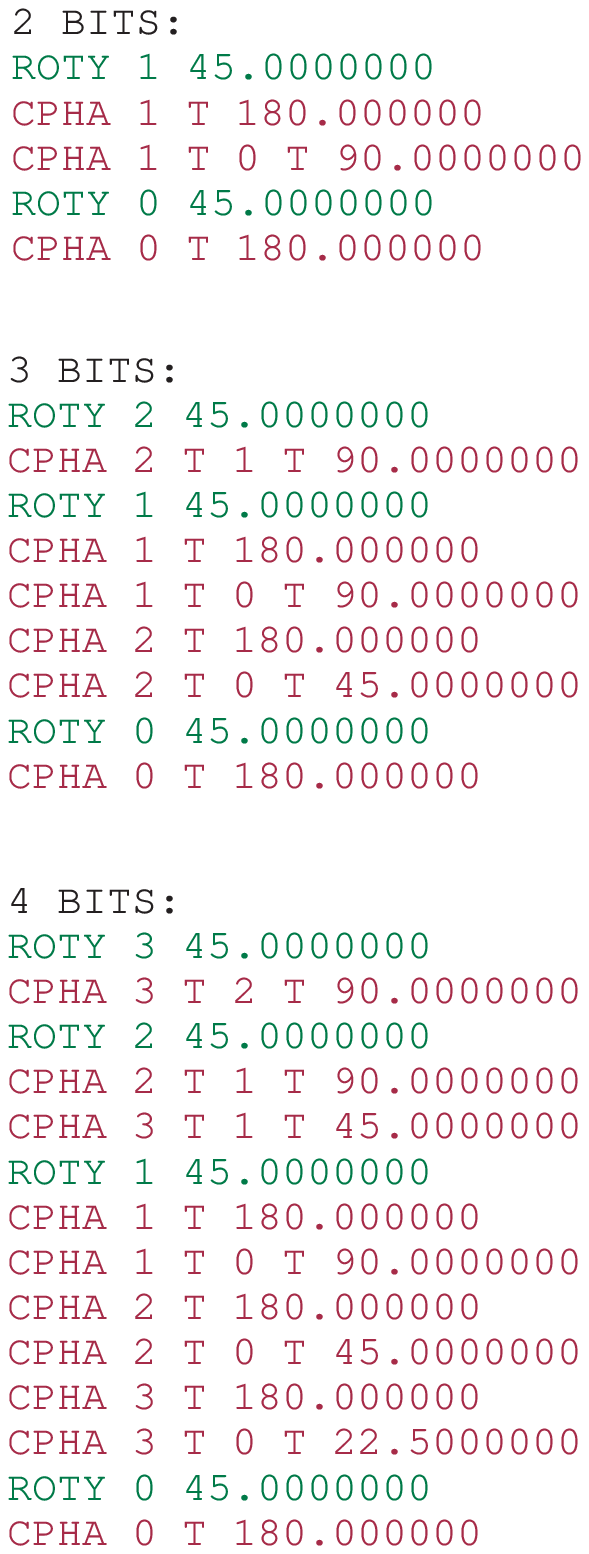}
			
			{Fig.10 Output of Qubiter with input matrix equal to 
			$P_{BR}F_{\nb}$ for $\nb = 2, 3, 4$.
			$F_{\nb}$ is the $\nb$ bit DFT matrix defined in Section 2(d).}
		\end{center}

\section*{Appendix A. We Can Assume CSD Angles Are\\
Non-decreasing Elements of $[0, 90^o]$}
\mbox{}\indent	
In this appendix we will show that: given any CSD of $U$
as in Eq.(1b.1),
one can always find
another CSD of $U$ for which the angles $\theta_i$ are in non-decreasing
order and they are contained in the interval $[0, 90^0]$.

For simplicity, suppose the number of CSD angles is two. The discussion 
that follows can be easily generalized to any number of angles.
For two CSD angles, Eq.(1b.1) becomes

\beq
U = 
\left [
\begin{array}{cc}
L_0 &  0 \\
0   &  L_1
\end{array}
\right ]
\left [
\begin{array}{cccc}
c_1		&0		&s_1	&0		\\
0  		&c_2 	&0		&s_2	\\
-s_1	&0 		&c_1	&0 		\\
0  		&-s_2 	&0		&c_2	
\end{array}
\right ]
\left [
\begin{array}{cc}
R_0 &  0 \\
0   &  R_1
\end{array}
\right ]
\;.
\eqlabel{A.1}\eeq

Suppose $c_2<0$. Then 

\beq
U = 
\left [
\begin{array}{cc}
L'_0 &  0 \\
0   &  L'_1
\end{array}
\right ]
\left [
\begin{array}{cccc}
c_1		&0		&s_1	&0		\\
0  		&|c_2| 	&0		&-s_2	\\
-s_1	&0 		&c_1	&0 		\\
0  		&s_2 	&0		&|c_2|	
\end{array}
\right ]
\left [
\begin{array}{cc}
R_0 &  0 \\
0   &  R_1
\end{array}
\right ]
\;,
\eqlabel{A.2a}\eeq
where

\beq
L'_j = L_j diag(1, -1)
\;,
\eqlabel{A.2b}\eeq
for $j=0, 1$. Hence, we may assume $c_1, c_2\geq 0$.

Suppose $c_1, c_2 \geq 0$ but $s_2<0$. Then

\beq
U = 
\left [
\begin{array}{cc}
L_0 &  0 \\
0   &  L'_1
\end{array}
\right ]
\left [
\begin{array}{cccc}
c_1		&0		&s_1	&0		\\
0  		&c_2 	&0		&|s_2|	\\
-s_1	&0 		&c_1	&0 		\\
0  		&-|s_2|	&0		&c_2	
\end{array}
\right ]
\left [
\begin{array}{cc}
R_0 &  0 \\
0   &  R'_1
\end{array}
\right ]
\;,
\eqlabel{A.3a}\eeq
where

\beq
L'_1 = L_1 diag(1, -1)
\;,
\eqlabel{A.3b}\eeq

\beq
R'_1 =  diag(1, -1)R_1
\;.
\eqlabel{A.3c}\eeq
Hence we may assume $c_1, c_2, s_1, s_2 \geq 0$.

Finally, suppose that $\theta_1> \theta_2$. Then

\beq
U = 
\left [
\begin{array}{cc}
L'_0 &  0 \\
0   &  L'_1
\end{array}
\right ]
\left [
\begin{array}{cccc}
c_2		&0		&s_2	&0		\\
0  		&c_1 	&0		&s_1	\\
-s_2	&0 		&c_2	&0 		\\
0  		&-s_1 	&0		&c_1	
\end{array}
\right ]
\left [
\begin{array}{cc}
R'_0 &  0 \\
0   &  R'_1
\end{array}
\right ]
\;,
\eqlabel{A.4a}\eeq
where

\beq
L'_j = L_j 
\left[
\begin{array}{cc}
0 & 1\\
1 & 0
\end{array}
\right]
\;,
\eqlabel{A.4b}\eeq

\beq
R'_j =  \left[
\begin{array}{cc}
0 & 1\\
1 & 0
\end{array}
\right]
R_j
\;,
\eqlabel{A.4c}\eeq
for $j=0,1$.
Hence, we may assume that $\theta_1 \leq \theta_2$.

\section*{Appendix B. Proof of Quantum FFT Algorithm}
\mbox{}\indent	
Without pretending to do justice to Coppersmith's beautiful
paper Ref.\enote{Copper}, here is a brief proof of his 
FFT algorithm, Eq.(2d.3), starting from the FFT algorithm that is usually 
found in textbooks. For simplicity, we will only consider 
the case $\nb =4$. What follows 
can be easily generalized to arbitrary $\nb\geq 1$. 

As usual, let

\beq
\omega = \exp(i \frac{2 \pi}{\ns} )
= \exp(i \frac{2 \pi}{2^4} ) 
\;.
\eqlabel{B.1}\eeq
Define matrices $\Omega, \Omega^2, \ldots$ by

\beq
\Omega = diag(1, \omega), \;\;\;
\Omega^2 = diag(1, \omega^2), 
\;\ldots
\eqlabel{B.2}\eeq
Prior to Refs.\enote{Copper}, the FFT algorithm might
have been stated like this: 
(see  Refs.\enote{Knuth} and \enote{Had})

\beq
F_4 = \frac{1}{\sqrt{2}}
(H\otimes I_8) 
[ I_8 \oplus 
(\Omega^4 \otimes \Omega^2 \otimes \Omega^1) ]
(I_2 \otimes F_3) P_4
\;,
\eqlabel{B.3}\eeq

\beq
F_3 = \frac{1}{\sqrt{2}}
(H\otimes I_4) 
[ I_4 \oplus 
(\Omega^4 \otimes \Omega^2) ]
(I_2 \otimes F_2) P_3
\;,
\eqlabel{B.4}\eeq

\beq
F_2 =\frac{1}{\sqrt{2}}
(H\otimes I_2) 
[ I_2 \oplus 
\Omega^4 ) ]
(I_2 \otimes F_1) P_2
\;,
\eqlabel{B.5}\eeq

\beq
F_1 = \frac{1}{\sqrt{2}}
H
\;.
\eqlabel{B.6}\eeq
The matrices $P_4, P_3, P_2$ are permutation
matrices to be specified later.
Although we could have 
written just a single equation that combined all four Eqs.(B.3) to (B.6), we 
have chosen not to do this, so as to make 
explicit the recursive nature of the beast.
Note

\beq
\Omega^4 \otimes \Omega^2 \otimes \Omega =
diag(
1 , \omega,\omega^2, \omega^3,
\omega^4,\omega^5, \omega^6, \omega^7 ) =
\omega^{2^2 n(2) + 2^1 n(1) + 2^0 n(0) }
\;.
\eqlabel{B.7}\eeq
The last equation becomes clear when one
realizes that
$2^2 n(2) + 2^1 n(1) + 2^0 n(0)$ operating on a state
$\ket{a_3, a_2, a_1, a_0}$ gives the 
binary expansion of $d(0, a_2, a_1, a_0)$.
Once Eq.(B.7) sinks into the old bean, it is only a 
short step to the realization that:

\beq 
I_8 \oplus (\Omega^4 \otimes \Omega^2 \otimes \Omega) = 
\omega^{n(3)[2^2 n(2) + 2^1 n(1) + 2^0 n(0)]}
= e^{i[
n(3)n(2)\phi_2 +
n(3)n(1)\phi_3 +
n(3)n(0)\phi_4 ] }
\;,
\eqlabel{B.8}\eeq

\beq 
I_4 \oplus (\Omega^4 \otimes \Omega^2) = 
(\omega^2)^{n(2)[2^1 n(1) + 2^0 n(0)]}
= e^{i[
n(2)n(1)\phi_2 +
n(2)n(0)\phi_3 ] }
\;,
\eqlabel{B.9}\eeq

\beq 
I_2 \oplus (\Omega^4) = 
(\omega^4)^{n(1)[2^0 n(0)]}
= e^{i[
n(1)n(0)\phi_2 ] }
\;.
\eqlabel{B.10}\eeq
The final step is to replace all
tensor products that contain H  by 
bit operators:

\beq
H\otimes I_8 = H(3)
\;,
\eqlabel{B.11}\eeq

\beq
I_2 \otimes H \otimes I_4 = H(2)
\;,
\eqlabel{B.12}\eeq

\beq
I_4 \otimes H \otimes I_2 = H(1)
\;,
\eqlabel{B.13}\eeq

\beq
I_8 \otimes H = H(0)
\;.
\eqlabel{B.14}\eeq

		\begin{center}
			\epsfig{file=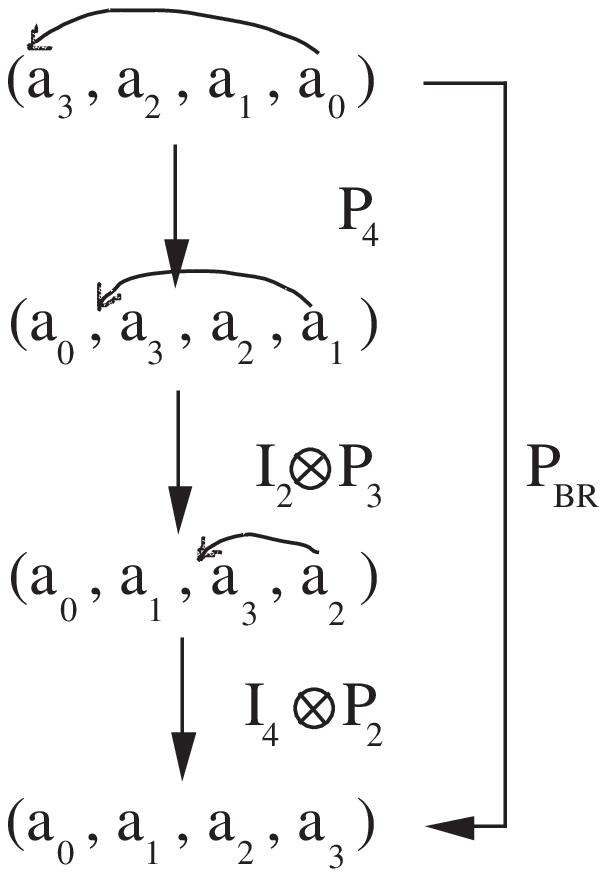}
			
			{Fig.11 Permutation matrices $P_4, P_3, P_2, P_{BR}$
			that arise in the FFT algorithm for $\nb=4$. }
		\end{center}


If all the permutation matrices of Eqs.(B.3) to (B.6) are
combined into a single permutation $P_{BR}$, then 

\beq
P_{BR} = (I_4 \otimes P_2)(I_2 \otimes P_3) P_4
\;.
\eqlabel{B.15}\eeq
Let $\va\in Bool^4$ label the columns of a $16\times 16$ matrix.
As was known previous to  Ref.\enote{Copper},
the matrices $P_4, P_3, P_2$ act as shown in Fig.11. Therefore,
their product $P_{BR}$ takes 
$\va = (a_3, a_2, a_1, a_0)$ to
$(a_0, a_1, a_2, a_3)$; i.e., $P_{BR}$ reverses the bits of $\va$.
$P_{BR}$ is not an elementary operation. However,
since it is a bit permutation,
it can easily be expressed as a product of Exchangers, which are elementary. 

\section*{Appendix C. Example of Decomposition \\of Controlled Gate into SEO}
\mbox{}\indent

Suppose $\nb \geq 3$ and consider the following controlled gate:

\beq
g = B(2)^{n(1)n(0)}
\;.
\eqlabel{C.1}\eeq
Replacing $n$ by $\frac{1}{2}[1-\sx]$, we get

\beq
g = B(2)^{\frac{1}{4} [1-\sz(1)] [1-\sz(0)]}
\;.
\eqlabel{C.2}\eeq
Expanding the exponent of $B(2)$ in Eq.(C.2), and 
using identity Eq.(4a.9), we get 

\beq
g = 
A_{00}
A_{01}
A_{11}
A_{10}
\;,
\eqlabel{C.3}\eeq
where

\beq
A_{00} = B(2)^{\frac{1}{4}}
\;,
\eqlabel{C.4}\eeq

\beq
A_{01} = B(2)^{\frac{-1}{4} \sz(0)}
= B(2)^{\frac{1}{4} n(0)} B(2)^{\frac{-1}{4}}
\;,
\eqlabel{C.5}\eeq

\beq
A_{11} = B(2)^{\frac{1}{4} \sz(1)\sz(0)}
=\cnotyes{0}{1}\odot B(2)^{\frac{1}{4} \sz(1)}
=[\cnotyes{0}{1}\odot B(2)^{\frac{-1}{4} n(1)}] B(2)^{\frac{1}{4}}
\;,
\eqlabel{C.6}\eeq

\beq
A_{10} = B(2)^{\frac{-1}{4} \sz(1)}
=B(2)^{\frac{1}{4} n(1)} B(2)^{\frac{-1}{4}}
\;.
\eqlabel{C.7}\eeq

The two $B(2)^{\frac{1}{4}}$ 
and two $B(2)^{\frac{-1}{4}}$ factors
cancel each other out.
We have ordered the factors $A_\vb$ in Eq.(C.3)
so that their subscripts $\vb\in Bool^2$ are in a lazy
ordering. (Ref.\enote{CastOfThousands} does this too.)
For the above case, lazy ordering is not important.
But when $g$ has more than two controls,
lazy ordering permits some c-nots between adjacent $A_\vb$'s to cancel.
The same techniques can be used
to decompose a controlled gate with more than two controls. 

\section*{Appendix D. Transforming between \\
${\cal B}(n, \nbar)$ and ${\cal B}(\sz, I_2)$ bases}
\mbox{}\indent	
In this appendix, we will show how to transform 
between the ${\cal B}(n, \nbar)$ and ${\cal B}(\sz, I_2)$
bases of the set of $\ns \times \ns$ diagonal complex  matrices.

Suppose that $\nb=1$ and let

\beq
X = 
\left[
\begin{array}{c}
\nbar(0) \\
n(0)
\end{array}
\right]
\;,
\eqlabel{D.1}\eeq

\beq
X' = 
\left[
\begin{array}{c}
1 \\
\sz(0)
\end{array}
\right]
\;.
\eqlabel{D.2}\eeq
Then

\beq
X = M_{1} X'
\;,
\eqlabel{D.3}\eeq
where

\beq
M_{1} = 
\frac{1}{2}
\left[
\begin{array}{rr}
1 & 1 \\
1 & -1
\end{array}
\right]
=
\frac{1}{2}
H_{1}
\;.
\eqlabel{D.4}\eeq
Note that

\beq
M_{1}^{-1} = H_{1}
\;.
\eqlabel{D.5}\eeq

Next suppose that $\nb = 2$. Let 

\beq
X = 
\left[
\begin{array}{c}
\nbar(1)\nbar(0) \\
\nbar(1)n(0) \\
n(1)\nbar(0) \\
n(1) n(0)
\end{array}
\right]
=
\left[
\begin{array}{c}
\nbar(1)\\
n(1)
\end{array}
\right]
\otimes 
\left[
\begin{array}{c}
\nbar(0)\\
n(0)
\end{array}
\right]
\;,
\eqlabel{D.6}\eeq

\beq
X' = 
\left[
\begin{array}{c}
1 \\
\sz(0)\\
\sz(1)\\
\sz(1)\sz(0)
\end{array}
\right]
=
\left[
\begin{array}{c}
1\\
\sz(1)
\end{array}
\right]
\otimes 
\left[
\begin{array}{c}
1\\
\sz(0)
\end{array}
\right]
\;.
\eqlabel{D.7}\eeq
Then

\beq
X = M_{2} X'
\;,
\eqlabel{D.8}\eeq
where

\beq
M_{2} =
\frac{1}{2^2}
\begin{tabular}{r|rrrr}
         & {\tiny 00} & {\tiny 01} & {\tiny 10} & {\tiny 11}\\
\hline
{\tiny 00}&  1&  1&  1&  1\\
{\tiny 01}&  1& -1&  1& -1\\
{\tiny 10}&  1&  1& -1& -1\\
{\tiny 11}&  1& -1& -1&  1\\
\end{tabular}
 =\frac{1}{2^2}
H_{1} \otimes H_{1}
=
\frac{1}{2^2}
H_{2}
\;.
\eqlabel{D.9}\eeq
Note that

\beq
M_{2}^{-1} = H_{2}
\;.
\eqlabel{D.10}\eeq

The above results for $\nb=1, 2$ can be easily generalized to any $\nb\geq 1$.
 One can prove using induction that for any $\nb$, 
the matrix $M_{\nb}$ has entries given by

\beq
(M_{\nb})_{a, b}  = \frac{1}{\ns}(-1)^{\vec{a}\cdot \vec{b}}
\;,
\eqlabel{D.11}\eeq
where $a, b \in Z_{0, \ns-1}$, $a = d(\vec{a})$ and  $b = d(\vec{b})$.
The matrix $M_{\nb}$ is just the $\nb$ bit Sylvester-Hadamard matrix
divided by $1/\ns$.

If $v$ and $v'$ are $\ns$ dimensional complex vectors such 
that 
\beq
v^T X = v'^T X'
\;,
\eqlabel{D.12}\eeq
then

\beq
X = M_{\nb} X'
\;,
\eqlabel{D.13}\eeq
implies

\beq
v' = M_{\nb}^T v
\;.
\eqlabel{D.14}\eeq

\section*{Appendix E. Transforming between \\
${\cal B}(n, \nbar)$ and ${\cal B}(n, I_2)$ bases}
\mbox{}\indent	
In this appendix, we will show how to transform 
between the ${\cal B}(n, \nbar)$ and ${\cal B}(\sz, I_2)$
bases of the set of $\ns \times \ns$ diagonal complex  matrices.

Suppose that $\nb=1$ and let

\beq
X = 
\left[
\begin{array}{c}
\nbar(0) \\
n(0)
\end{array}
\right]
\;,
\eqlabel{E.1}\eeq

\beq
X' = 
\left[
\begin{array}{c}
1 \\
n(0)
\end{array}
\right]
\;.
\eqlabel{E.2}\eeq
Then

\beq
X = M_{1} X'
\;,
\eqlabel{E.3}\eeq
where

\beq
M_{1} = 
\left[
\begin{array}{rr}
1 & -1 \\
0 & 1
\end{array}
\right]
\;.
\eqlabel{E.4}\eeq
Note that

\beq
M_{1}^{-1} = 
\left[
\begin{array}{cc}
1 & 1 \\
0 & 1
\end{array}
\right]
\;.
\eqlabel{E.5}\eeq

Next suppose that $\nb = 2$. Let 

\beq
X = 
\left[
\begin{array}{c}
\nbar(1)\nbar(0) \\
\nbar(1)n(0) \\
n(1)\nbar(0) \\
n(1) n(0)
\end{array}
\right]
=
\left[
\begin{array}{c}
\nbar(1)\\
n(1)
\end{array}
\right]
\otimes 
\left[
\begin{array}{c}
\nbar(0)\\
n(0)
\end{array}
\right]
\;,
\eqlabel{E.6}\eeq

\beq
X' = 
\left[
\begin{array}{c}
1 \\
n(0)\\
n(1)\\
n(1)n(0)
\end{array}
\right]
=
\left[
\begin{array}{c}
1\\
n(1)
\end{array}
\right]
\otimes 
\left[
\begin{array}{c}
1\\
n(0)
\end{array}
\right]
\;.
\eqlabel{E.7}\eeq
Then

\beq
X = M_{2} X'
\;,
\eqlabel{E.8}\eeq
where

\beq
M_{2} =
\begin{tabular}{r|rrrr}
         & {\tiny 00} & {\tiny 01} & {\tiny 10} & {\tiny 11}\\
\hline
{\tiny 00}&  1 &-1 &-1 & 1 \\
{\tiny 01}&  0 & 1 & 0 &-1 \\
{\tiny 10}&  0 & 0 & 1 &-1 \\
{\tiny 11}&  0 & 0 & 0 & 1
\end{tabular}
= M_{1} \otimes M_{1}
\;.
\eqlabel{E.9}\eeq
Note that

\beq
M_{2}^{-1} = M_{1}^{-1} \otimes M_{1}^{-1}
\;.
\eqlabel{E.10}\eeq

For any non-negative integers $a, b$, let
$a \& b$ and $a \wedge b$ denote the same thing as they do in 
the C and C++ computer programming languages. That is,
$a \& b$ is the result of bitwise and-ing of $a$ and $b$,
and $a \wedge b$ is the result of bitwise xor-ing of $a$ and $b$
(which is the same as bitwise mod(2) addition of $a$ and $b$).
For any $\va \in Bool^\nb$, if $a = d(\va)$, we define
$||a||_1 = ||\va||_1 = \sum_{\mu=0}^{\nb-1} a_\mu$. In other words,
$||a||_1$ is the number of ON bits in $\va$.

The above results for $\nb=1, 2$ can be easily generalized to any $\nb\geq 1$.
One can prove using induction that for any $\nb$, 
the matrix $M_{\nb}$ has entries given by

\beq
(M_{\nb})_{a, b}  = (-1)^{||a\wedge b||_1} \delta(a \& b, a)
\;,
\eqlabel{E.11}\eeq
where $a, b \in Z_{0, \ns-1}$.

\section*{Appendix F. Central Matrices\\
Whose Entries Are Elements of $\{-1, 0, 1\}$}
\mbox{}\indent	
In this paper, we use bare central matrices whose D matrices have angles
contained in the interval $[0, 90^o]$.
An interesting special case is when these angles are all
either 0 or $90^o$ exclusively. 
The entries of such central matrices are elements
of $\{-1, 0, 1\}$. 

First suppose that the central matrix $Q$ is
a single D matrix. Then we can express
$Q$ as

\beq
Q = \exp \left( i 
\sum_{\vb \in Bool^{\nb-1}}
\phi_\vb 
\sigma_y \otimes P_\vb
\right)
\;,
\eqlabel{F.1}\eeq
where the $\phi_b$ are all either 0 or $90^o$.
Thus

\beq
Q = [e^{i \frac{\pi}{2} \sigma_y(\nb - 1)}]^{\sum_{\vb\in S} P_\vb} 
= \prod_{\vb\in S} [e^{i \frac{\pi}{2} \sigma_y(\nb - 1)}]^{P_\vb}
\;,
\eqlabel{F.2a}\eeq
where

\beq
S = \{\vb| \vb\in Bool^{\nb-1},\;\; \phi_\vb = \frac{\pi}{2} \}
\;.
\eqlabel{F.2b}\eeq
Hence, $Q$ can always be expressed as a product
of controlled qubit rotations. 
If $X = \sum_{\vb\in S} P_\vb$, then 
$Q$ always acts
on more than 2 bits at a time except when 
(1)$X=1$ or when (2)$X=P_b(\alpha)$
for some $b\in Bool$ and some $\alpha\in\bitz$ such that $\alpha \neq \nb-1$.
In the first case, $Q$ is just a simple qubit rotation. In the second case 

\beq
Q =  [e^{i \frac{\pi}{2} \sigma_y(\nb - 1)}]^{P_b(\alpha)}
\;.
\eqlabel{F.3}\eeq

Next suppose that $Q$ is a direct sum of several D matrices.
As explained in Section 4(b), such a $Q$ can be obtained 
by applying a bit permutation matrix to a $Q$ that, 
like the one in Eq.(F.1), is a single D matrix.

Thus, when $Q$ is a direct sum of one or more D matrices,
Q always acts on more than 2 bits except when 
it is a simple qubit rotation or when it has the form

\beq
Q =  [e^{i \frac{\pi}{2} \sigma_y(\gamma)}]^{P_b(\alpha)}
\;,
\eqlabel{F.4}\eeq
where $\alpha$ and $\gamma$ are distinct elements of $\bitz$
and $b\in Bool$. Below, we will show that Eq.(F.4) implies

\beq
Q
=
\sigma_x(\gamma)^{P_b(\alpha)} (-1)^{\nbar(\gamma) P_b(\alpha)}
=
 (-1)^{n(\gamma) P_b(\alpha)} \sigma_x(\gamma)^{P_b(\alpha)}
\;.
\eqlabel{F.5}\eeq
Hence, such a $Q$ can be expressed as a product of a c-not and a controlled
phase shift with 2 controls.
Whenever the optimization of Section 5(c)
(Extracting Phases from Complex D Matrices) is ON, we will
decompose any $Q$ of the form Eq.(F.4)
into the SEO of Eq.(F.5).

Note that 

\beq
e^{i \frac{\pi}{2} \sigma_y} =
\left[
\begin{array}{cc}
0  & 1 \\
-1 & 0 
\end{array}
\right]
\;,
\eqlabel{F.6}\eeq

\beq
(-1)^{\nbar} = diag(-1, 1)
\;,
\eqlabel{F.7}\eeq

\beq
(-1)^{n} = diag(1,-1)
\;.
\eqlabel{F.8}\eeq
Thus,
 
\beq
e^{i \frac{\pi}{2} \sigma_y} =
\sigma_x (-1)^{\nbar} =
(-1)^n \sigma_x
\;.
\eqlabel{F.9}\eeq
($n$ changes to $\nbar$ because 
$\sigma_x \sigma_z = -\sigma_z \sigma_x$.)
Eq.(F.5) follows directly from Eq.(F.9).

Note that we could also decompose the $Q$
of Eq.(F.4) using the techniques of Section 4 and Appendix C:

\beq
\begin{array}{ll}
e^{i \frac{\pi}{2} \sigma_y(\gamma) n(\alpha)} &
 = e^{i \frac{\pi}{4} \sigma_y(\gamma)}e^{-i \frac{\pi}{4} \sigma_y(\gamma)\sigma_z(\alpha)}\\
 &
 =[e^{i \frac{\pi}{4} \sigma_y(\gamma)}]
 [\cnotyes{\alpha}{\gamma}\odot
 e^{-i \frac{\pi}{4} \sigma_y(\gamma)}]
 \end{array}
 \;.
 \eqlabel{F.10}\eeq
 Neither decomposition Eq.(F.5) nor decomposition Eq.(F.10)
 uses controlled qubit rotations. 
 The SEO of Eq.(F.5) is shorter than the SEO of Eq.(F.10).
 Eq.(F.5)
 uses a controlled phase shift whereas 
Eq.(F.10) doesn't. Since controlled phase shifts are
common when the optimization of Section 5(c)
is ON, it is natural to use Eq.(F.5) when said optimization is ON.

\newpage
\section*{FIGURE CAPTIONS:}
\begin{description}

\item{\sc Fig.1} A CSD binary tree. 

\item{\sc Fig.2} Pictorial representation of
quantum FFT algorithm. For simplicity, we
only show the Hadamard matrices.
White (unshaded) regions inside a matrix
represent zero entries. 

\item{\sc Fig.3} Circuit symbols for the 4 different types of c-nots.

\item{\sc Fig.4} Four equivalent circuit diagrams for Exchanger.

\item{\sc Fig.5}  Four equivalent circuit diagrams for Twin-to-twin-er.

\item{\sc Fig.6} Circuit symbol for Exchanger.

\item{\sc Fig.7} Circuit diagram for Eq.(3b.7).

\item{\sc Fig.8} Subset of a CSD tree.

\item{\sc Fig.9} Output of Qubiter with input matrix equal to 
$\frac{1}{\sqrt{\ns}} H_{\nb}$ for $\nb = 2, 3, 4$.
$H_{\nb}$ is the $\nb$ bit Hadamard matrix defined in Section 2(c).
	
\item{\sc Fig.10} Output of Qubiter with input matrix equal to 
$P_{BR}F_{\nb}$ for $\nb = 2, 3, 4$.
$F_{\nb}$ is the $\nb$ bit DFT matrix defined in Section 2(d).

\item{\sc Fig.11} Permutation matrices $P_4, P_3, P_2, P_{BR}$
that arise in the FFT algorithm for $\nb=4$.
	
\end{description}

\end{document}